\documentclass[fleqn,usenatbib]{mnras}

\usepackage{newtxtext,newtxmath}

\usepackage[T1]{fontenc}
\usepackage{ae,aecompl}

\usepackage{graphicx}	
\usepackage{amsmath}	
\usepackage{amssymb}	
\usepackage{placeins}

\usepackage{etoolbox}
\makeatletter
\patchcmd\@combinedblfloats{\box\@outputbox}{\unvbox\@outputbox}{}{\errmessage{\noexpand patch failed}}
\makeatother

\title[KGES]{From Peculiar Morphologies to Hubble--type Spirals: The relation between galaxy dynamics and morphology in star--forming galaxies at $z$\,$\sim$\,1.5}

\author[S. Gillman et al.]{S. Gillman,$^{1}$\thanks{E-mail: steven.r.gillman@durham.ac.uk}
A. L.  Tiley,$^{1,2}$ 
A. M. Swinbank,$^{1}$
C. M. Harrison,$^{3}$
Ian Smail,$^{1}$ 
\and U. Dudzevi\v{c}i\={u}t\.{e},$^{1}$
R. M. Sharples,$^{1,4}$
L. Cortese,$^{2,5}$
D. Obreschkow,$^{2,6}$
R. G. Bower$^{1,7}$
\and
T. Theuns$^{7}$
M. Cirasuolo,$^{3}$
D. B. Fisher,$^{8}$
K. Glazebrook,$^{8}$
Edo Ibar,$^{9}$
\and 
J. Trevor Mendel,$^{10}$
and Sarah M. Sweet,$^{8,5}$ 
\\
$^{1}$Centre for Extragalactic Astronomy, Durham University, South Road, Durham, DH1 3LE UK\\
$^{2}$International Centre for Radio Astronomy Research, University of Western Australia, 35 Stirling Highway, Crawley, WA, Australia\\
$^{3}$European Southern Observatory, Karl-Schwarzschild-Str. 2, 85748 Garching b. M\"unchen, Germany \\
$^{4}$Centre for Advanced Instrumentation, Durham University, South Road, Durham DH1 3LE UK\\
$^{5}$ARC Centre of Excellence for All Sky Astrophysics in 3 Dimensions (ASTRO 3D) \\
$^{6}$Australian Research Council Centre of Excellence for All-Sky Astrophysics, 44 Rosehill Street Redfern, NSW 2016, Australia \\
$^{7}$Institute for Computational Cosmology, Durham University, South Road, Durham DH1 3LE UK\\
$^{8}$Centre for Astrophysics and Supercomputing, Swinburne University of Technology, PO Box 218, Hawthorn, VIC 3122, Australia \\
$^{9}$Instituto de F\'isica y Astronom\'ia, Universidad de Valpara\'iso, Avda. Gran Breta\~na 1111, Valpara\'iso, Chile\\
$^{10}$Research School of Astronomy and Astrophysics, Australian National University, Canberra, ACT 2611, Australia \\
}

\date{Accepted 2019 December 18. Received 2019 December 13; in original form 2019 October 8}

\pubyear{2019}

\begin{document}
\label{firstpage}
\pagerange{\pageref{firstpage}--25}
\maketitle

\begin{abstract}
We present an analysis of the gas dynamics of star--forming galaxies at $z$\,$\sim$\,1.5 using data from the KMOS Galaxy Evolution Survey (KGES). We quantify the morphology of the galaxies using $HST$ {\sc{CANDELS}} imaging  parametrically and non-parametrically. We combine the  H$\alpha$ dynamics from KMOS with the high--resolution imaging to derive the relation between stellar mass (M$_{*}$) and stellar specific angular momentum (j$_{*}$). We show that high--redshift star--forming galaxies at $z$\,$\sim$\,1.5 follow a  power-law trend in specific stellar angular momentum with stellar mass similar to that of local late--type galaxies of the form j$_*$\,$\propto$\,M$_*^{0.53\,\pm\,0.10}$. The highest specific angular momentum galaxies are mostly disc--like, although generally, both peculiar morphologies and disc-like systems are found across the sequence of specific angular momentum at a fixed stellar mass.  We explore the scatter within the j$_{*}$\,--\,M$_{*}$ plane and its correlation with both the integrated dynamical properties of a galaxy (e.g. velocity dispersion, Toomre Q$_{\rm g}$, H$\alpha$ star formation rate surface density $\Sigma_{\rm SFR}$) and its parameterised rest-frame UV\,/\,optical morphology (e.g. S\'ersic index, bulge to total ratio, Clumpiness, Asymmetry and Concentration). We establish that the position in the j$_{*}$\,--\,M$_{*}$ plane is correlated with the star-formation surface density and the Clumpiness of the stellar light distribution. 
Galaxies with peculiar rest-frame UV\,/\,optical morphologies have comparable specific angular momentum to disc\,--\,dominated galaxies of the same stellar mass, but are clumpier and have higher star-formation rate surface densities.
We propose that the peculiar morphologies in high--redshift systems are driven by higher star formation rate surface densities and higher gas fractions leading to a more clumpy inter-stellar medium.

\end{abstract}




\section{Introduction}
In 1926, Edwin Hubble established the Hubble-Sequence of galaxy morphology by visually classifying local galaxies into distinct classes of spirals, ellipticals, lenticulars and peculiars \citep{Hubble1926}. The Hubble-Sequence 
remains one of the defining
characteristics of galaxies, and provides one of the key constraints
that galaxy formation models strive to reproduce (e.g. \citealt{Tissera1990}; \citealt{Synder2015};  \citealt{Trayford2018}, \citealt{Zoldan2019}.)
As originally suggested by \citet{Sandage1970}, dynamical surveys of local galaxies suggest that the Hubble-Sequence of galaxy morphologies follows a sequence of increasing angular momentum at a fixed mass (e.g. \citealt{Sandage1986,Hernandez2006,Hammer2009,Falcon2015})

In the cold dark matter paradigm, galaxies form at the centres of dark matter halos.  As the dark matter halos grow early in their formation history, they acquire angular momentum (J) as a result
of large-scale tidal torques that arise from the growth of perturbations \citep{Stewart2017}.  The specific angular momentum acquired has a strong mass dependence, with j\,$\propto$\,M$_{\rm halo}^{2/3}$
(e.g.\ \citealt{Catelan1996}).  As the gas collapses within the
halo from the virial radius to the disc scale, the baryons can both
lose and gain angular momentum. The models suggest that late--type galaxies (e.g. star--forming, discy, dynamically young systems), are those that better preserve the halo dynamical properties. The (weak) conservation of baryonic angular momentum during collapse results 
in a centrifugally supported disc with an exponential mass profile (e.g. \citealt{Mo1998}).
Early--type galaxies, in contrast, have either a very low retention factor of the baryonic angular momentum, (e.g. \citealt{Onghia2006,Sokolowska2017}) or reside in dark matter halos with low spin, likely due to mergers and disc instabilities (e.g. \citealt{Hernandez2007,Gomez2017}).

\citet{Fall1980} established that the specific stellar angular momentum, j$_*$\,=\,J/M$_*$, of low redshift massive disc galaxies follows a tight sequence with stellar mass quantified as j$_*$\,$\propto$\,M$_*^{2/3}$. This  j$_*$--M$_*$ plane was shown by  \citet{Romanowsky2012} to correlate with galaxy morphology, with early--type galaxies having a factor of $\sim$5\,$\times$ less specific angular momentum than late--type galaxies of the  same stellar mass. More recent integral field studies of low redshift galaxies have analysed the connection between a galaxy's parameterised morphology (e.g. S\'ersic index, stellar bulge to total ratio) and specific angular momentum (\citealt{Glazebrook2014}; \citealt{Cortese2016}). More bulge dominated galaxies, with higher S\'ersic indices, have been shown to have lower specific angular momentum at fixed stellar mass (\citealt{Fall2018}). The scatter about the j$_*$\,$\propto$\,M$_*^{2/3}$ sequence in the local Universe is driven by the variation in the combination of disc and bulge components that make up star--forming late--type galaxies at $z$\,$\sim$\,0
(e.g. \citealt{Romeo2018}; \citealt{Sweet2018};  \citealt{Jadhav2019}).
 

While the role of angular momentum in locating galaxies along the
Hubble-Sequence is well constrained at $z$\,$\sim$\,0, the relationship 
between  angular momentum and the emergence of the Hubble-Sequence at high redshift
is less established. Early work by  \citet{Puech2007} established that star--forming galaxies at intermediate redshifts ($z$\,$\sim$\,0.6) have comparable dynamical properties to local galaxies. Galaxies identified to have complex kinematics however, exhibit significantly more scatter in dynamical scaling relations, with higher levels of turbulence indicating the presence of mergers and interactions.
At higher redshift, morphological and dynamical studies have
shown that the high-redshift ($z$\,$\sim$\,2) star--forming  galaxy population is
dominated by turbulent, gas-rich systems
\citep[e.g.][]{Bouche2007,Genzel2011,Wisnioski2015}. 
Multi-wavelength imaging has been used to  identify a transformation in galaxy morphology from 
single component systems (bulge or disc) to two component (bulge and disc) systems
around $z$\,$\sim$\,2 (e.g. \citealt{Sachdeva2019}).
The transition in morphology is reflected in
other galaxy properties such as star formation, colour and stellar mass, indicating there is a wider 
physical mechanism responsible for the galaxies' evolution (e.g. \citealt{Bruce2014,Lang2014,Huertas-Company2015}).
The transition from a population dominated by clumpy, irregular morphologies to
morphologically smooth, disc-like galaxies appears to occur around $z$\,$\sim$\,1.5.  This epoch has therefore been heralded as the epoch when the Hubble-Sequence ``emerged'' (e.g. \citealt{Cowie1995}; \citealt{Conselice2011}). 

Numerical simulations, which attempt to model the galaxies across cosmic time, suggest that the  transition from galaxies with clumpy, irregular visual morphologies to well defined Hubble-like morphologies is also dependent on the strength and efficiency of feedback
controlling  star formation (e.g. \citealt{Benson2003,Okamoto2005,Sales2010}). 
The stellar mass and  specific angular momentum of the galactic disc grows as a consequence of the ongoing feedback and cosmological accretion, 
such that the disc is stable against large scale collapse 
(\citealt{Bournaud2014}; \citealt{Oklopcic2017}).  
In particular, the Evolution and Assembly of GaLaxies and their Environments ({\sc{eagle}}; \citealt{Crain2015}, \citealt{Schaye2015}) hydrodynamic simulation
has suggested that the morphology of galaxies of all masses at high--redshift are asymmetric, 
with a causal relationship between the morphology of a galaxy and its
host dark matter halo (e.g. \citealt{Trayford2018}; \citealt{Thob2019}). The 
scatter in the angular momentum of the baryons and stars within the {\sc{eagle}}
simulation correlates strongly with other galaxy properties 
such as, gas fraction, stellar concentration and the ratio of circular velocity
to velocity dispersion \citep{Lagos2017}. Recent semi-analytical models (SAMs) 
have further identified the relation between stellar and halo specific 
angular momentum exhibiting no redshift evolution, (e.g. \citealt{Marshall2019}), whilst the relationship between specific angular momentum and stellar mass increases by 0.5\,dex from
$z$\,=\,7 to $z$\,=\,2, with the dominant morphological fraction of high--redshift 
galaxies being bulge--dominated systems (e.g. \citealt{Zoldan2018,Zoldan2019,Tachella2019}).

Other high--resolution hydrodynamical zoom-in simulations, such as Feedback in Realistic Environments (FIRE; \citealt{Hopkins2014,Hopkins2018}), have shown that the stellar morphology and kinematics of Milky Way mass galaxies at low redshift correlate strongly with the gaseous history of the galaxy and less with the dark matter halo properties. In these simulations the likelihood of the formation of a well--ordered stellar discs below $z$\,$\sim$\,1 depends on the gas mass within the disc (e.g. \citealt{Garrison-Kimmel2018}) as well as the angular momentum of the system (e.g. \citealt{Obreschkow2016,Badry2018})

Most of the measurements of the internal dynamics of galaxies at this epoch, 
which are needed to test these models, have come from moderately small samples of a few tens of galaxies (e.g. \citealt{Schreiber2006}, \citealt{Contini2016}, \citealt{Posti2018}), making if difficult to constrain 
the physical processes driving the evolution in galaxy dynamics. Larger samples of high--redshift star--forming galaxy dynamics are becoming more available due to the next generation of extragalactic integral field surveys. For example, the KMOS$^{\rm 3D}$ survey \citep{Wisnioski2015} of $\sim$\,360 star--forming galaxies at $z$\,$\sim$\,1\,--\,3 established that the specific angular momentum of a disc galaxy reflects that of its host dark matter halo with the presence of a j$_*$--M$_*$ plane at this epoch \citep{Burkert2016}. By analysing the H$\alpha$ gas dynamics of $\sim$\,700 star--forming galaxies from the KMOS Redshift One Spectroscopic Survey (KROSS; \citealt{Stott2016}), \citet{Harrison2017} showed that the normalisation of the j$_*$--M$_*$ plane at $z$\,$\sim$\,1 was 0.2\,--\,0.3 dex lower compared to that of $z$\,$\sim$\,0 disc galaxies, indicating that high-redshift galaxies, at fixed stellar mass, have lower specific stellar angular momentum. It should be noted however that \citet{Marasco2019} concluded that there is no evolution in j$_*$--M$_*$ plane from $z$\,=\,0, in a small selected sample of $z$\,=\,1 disc galaxies.

The connection between galaxy morphology and the distribution of angular momentum  at $z$\,$\sim$\,0.5\,--\,1.5 was qualitatively established by \citet{Swinbank2017}, showing that galaxies with `visually' more disc dominated morphologies had higher angular momentum at fixed stellar mass whilst lower angular momentum galaxies had more peculiar `complicated' morphologies. This relationship was quantified further by \citet{Harrison2017}, who parameterised the morphology of the KROSS galaxies with S\'ersic profiles, establishing a trend of decreasing specific angular momentum, at fixed stellar mass, with increasing S\'ersic index, suggesting  there is a causal connection between morphology and angular momentum. Merger events and interactions also enhance gas velocity dispersion and reduce a galaxy's angular momentum, introducing significant scatter into dynamical scaling relations \citep[e.g.][]{Puech2019}

In order to quantify how the angular momentum of high--redshift star--forming galaxies affects the emergence of the Hubble--type disc galaxies, and the role feedback plays in defining a galaxy's morphology, we require two key quantities. First, we need to derive the internal dynamics and second, we need to measure 
rest--frame optical morphology of the galaxies at this epoch both, parametrically and non-parametrically, which requires high resolution multi--wavelength imaging of the galaxies.

In this paper we present and analyse the relation between gas dynamics, angular momentum and rest-frame optical morphology in a sample of 235 mass selected star--forming galaxies in the redshift range $z$\,=\,1.22\,--\,1.76. This survey, the KMOS Galaxy Evolution Survey (KGES; Tiley et. al. in prep.), represents a 27-night guaranteed time programme using the $K$-band Multi Object Spectrograph (KMOS; \citealt{Sharples2013}) which primarily targets star--forming galaxies in the $HST$ Cosmic Assembly Near-infrared Deep Extragalactic Legacy Survey (CANDELS; \citealt{Koekemoer2011}) with multi-wavelength imaging. We present the seeing-limited resolved H$\alpha$ dynamics of 235 galaxies, across a broad range of stellar mass and H$\alpha$ star formation rate, from which we measure each galaxys' dynamics and morphology. We analyse the connection between a galaxy's rest--frame optical morphology, quantified both parametrically and non-parametrically, and its fundamental dynamical properties that define the emergence of the Hubble-Sequence at $z$\,$\sim$1.5.

In Section \ref{Sec:Sample} we discuss the sample selection, observations and data reduction of the KMOS observations that make up the KGES Survey. In Section \ref{Sec:Analysis} we derive the galaxy integrated photometric and morphological properties, e.g. star formation rates, stellar mass, S\'ersic index and stellar continuum sizes. We then use the stellar continuum sizes and inclinations to derive the dynamical properties of the galaxies before combining the galaxy sizes, stellar masses and dynamical properties to measure the specific angular momentum of the KGES galaxies. In Section \ref{Sec:Diss} we discuss and interpret our findings, exploring the connection between galaxy morphology and dynamics, before giving our conclusions in Section \ref{Sec:Conc}.

A Nine-Year Wilkinson Microwave Anisotropy Probe \citep{Hinshaw2013} cosmology is used throughout this work with $\Omega_{\Lambda}$\,=\,0.721, $\Omega_{\rm m}$\,=\,0.279 and H$_{\rm 0}$\,=\,70\,km\,s$^{-1}$ Mpc$^{-1}$.
In this cosmology a spatial resolution of 0.65 arcsecond (the median FWHM of the seeing in our data) corresponds to a physical scale of 5.6\,kpc at a redshift of $z$\,=\,1.5. All quoted magnitudes are on the AB system and stellar masses are calculated assuming a Chabrier initial mass function (IMF) \citep{Chabrier2003}.

\section{Sample Selection, Observations and Data Reduction}\label{Sec:Sample}

\begin{figure}
	\centering
	\includegraphics[width=\linewidth]{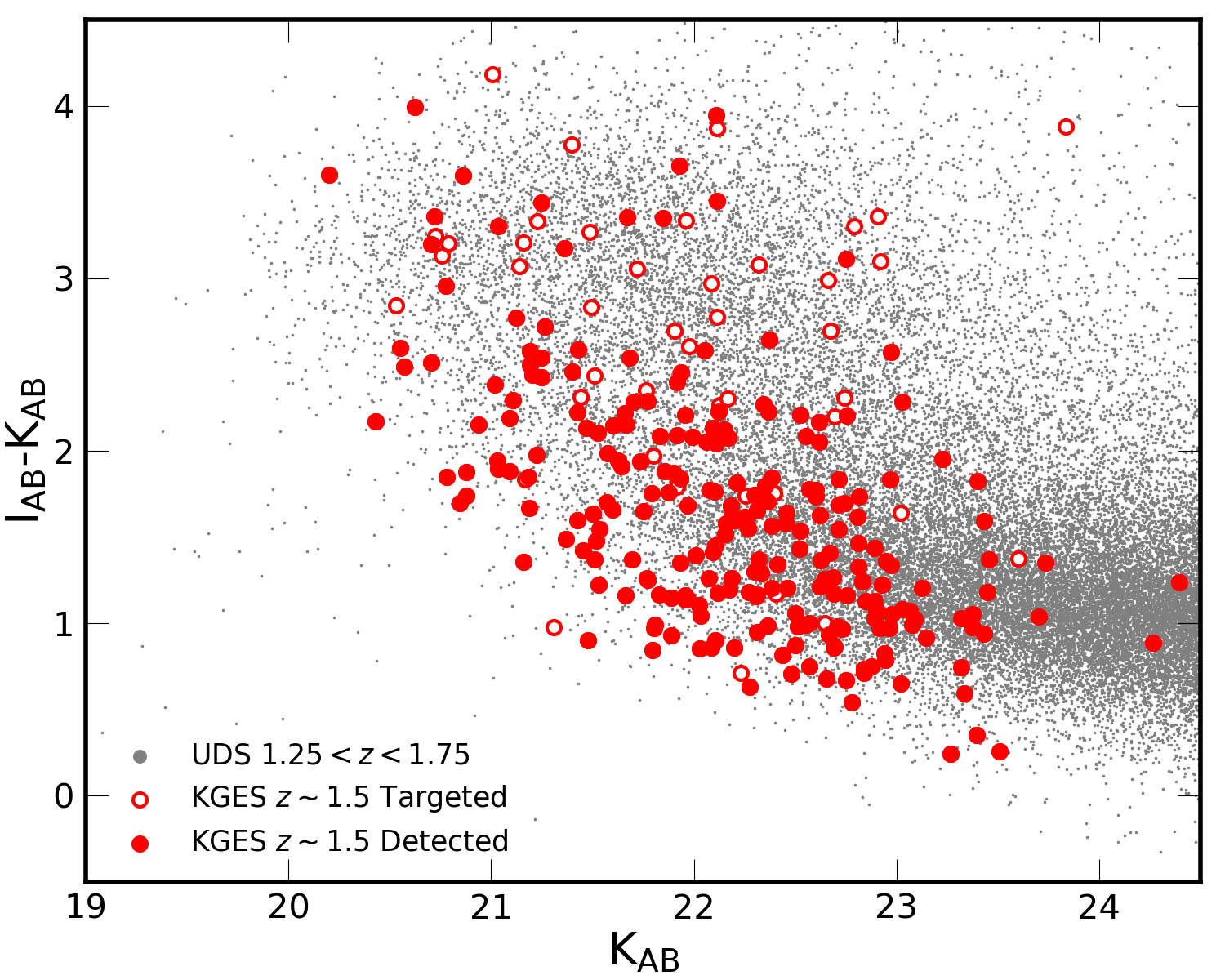}
	\caption{The observed ($I_{\rm AB}$\,--\,$K_{\rm AB}$) colour as a function of the observed $K$-band magnitude for the KGES sample. Galaxies detected in H$\alpha$ are indicated by the red points (243 galaxies). Open symbols represent the 45 galaxies where the H$\alpha$ signal to noise (S/N) is less than five. Star--forming galaxies in the UDS field in the redshift range 1.25\,$<$\,$z$\,$<$\,1.75 are shown for comparison (grey points).}
\label{Fig:MK}
\end{figure}

The KMOS Galaxy Evolution Survey (Tiley et. al. in prep.) concentrates on measuring the dynamics of `main--sequence' star--forming galaxies at $z$\,$\sim$\,1.5, and builds upon previous high--redshift surveys of star--forming galaxies \citep[e.g KROSS at $z$\,$\sim$\,0.9,][]{Stott2016, Harrison2017}. We predominately target  galaxies at $z$\,$\sim$\,1.5 in the $HST$ CANDELS field within the spectral range containing the redshifted H$\alpha$ $\lambda$6563$\AA$ and [N{\sc{ii}}] ($\lambda$6548, $\lambda$6583) nebular emission line to obtain a measure of the galaxies' ongoing star formation. The majority of galaxies in the KGES survey are selected to have  known spectroscopic redshifts and a $K$\,--\,band magnitude of $K$\,$<$\,22.5. If not enough galaxies pass this criteria to fill the KMOS arms in each mask, fainter galaxies were selected. We note that there was no morphological selection when selecting galaxies to be observed with KMOS. In Figure \ref{Fig:MK} we show an $I$\,--\,$K$ colour magnitude diagram for targeted and H$\alpha$ detected KGES galaxies. The galaxies in the survey occupy a similar region of colour magnitude parameter space to typical star--forming galaxies in the  UKIDSS Ultra-Deep Survey (UDS; \citealt{Lawrence2007}) field from $z$\,=\,1.25\,--\,1.75.

A full description of the survey design, observations and data reduction is presented in Tiley et al. (in prep.). In brief, we observed 288 high-redshift galaxies with KMOS as part of the KGES survey between October 2016 and January 2018. Each target was observed in five observing blocks (OB) for a total exposure time of 27ks in an ABAABA sequence (A\,=\,Object frame, B\,=\,Sky frame) with individual exposures of 600s. The median FWHM of the seeing in our observations is $\langle$\,FWHM\,$\rangle$\,=\,0.65\,$\pm$\,0.11 arcseconds with a range from FWHM\,=\,0.49\,--\,0.82 arcseconds. Our targets lie in the UDS, Cosmological Evolution Survey (COSMOS; \citealt{Scoville2007}) and  Extended Chandra Deep Field South (ECDFS; \citealt{Giacconi2001}) extragalactic fields.

The European Southern Observatory (ESO) Recipe Execution Tool ({\sc{ESOREX}}; \citealt{ESO2015}) pipeline was used to extract, wavelength calibrate and flat field each of the spectra and form a data cube from each observation. The sky-subtraction for the KGES observations is performed on a frame by frame basis, with an initial A--B subtraction. Before stacking, we employ the Zurich Atmospheric Purge ({\sc{zap}}; \citealt{Soto2016}) tool, adapted for use with KMOS,  which uses a principal component analysis to characterise and remove the remaining sky residuals in the observations (Mendel et al. in prep.). ZAP is trained on  residual sky spectra devoid of source emission derived from a median of the A--B frames.

\begin{figure*}
	\centering
	\includegraphics[width=\linewidth]{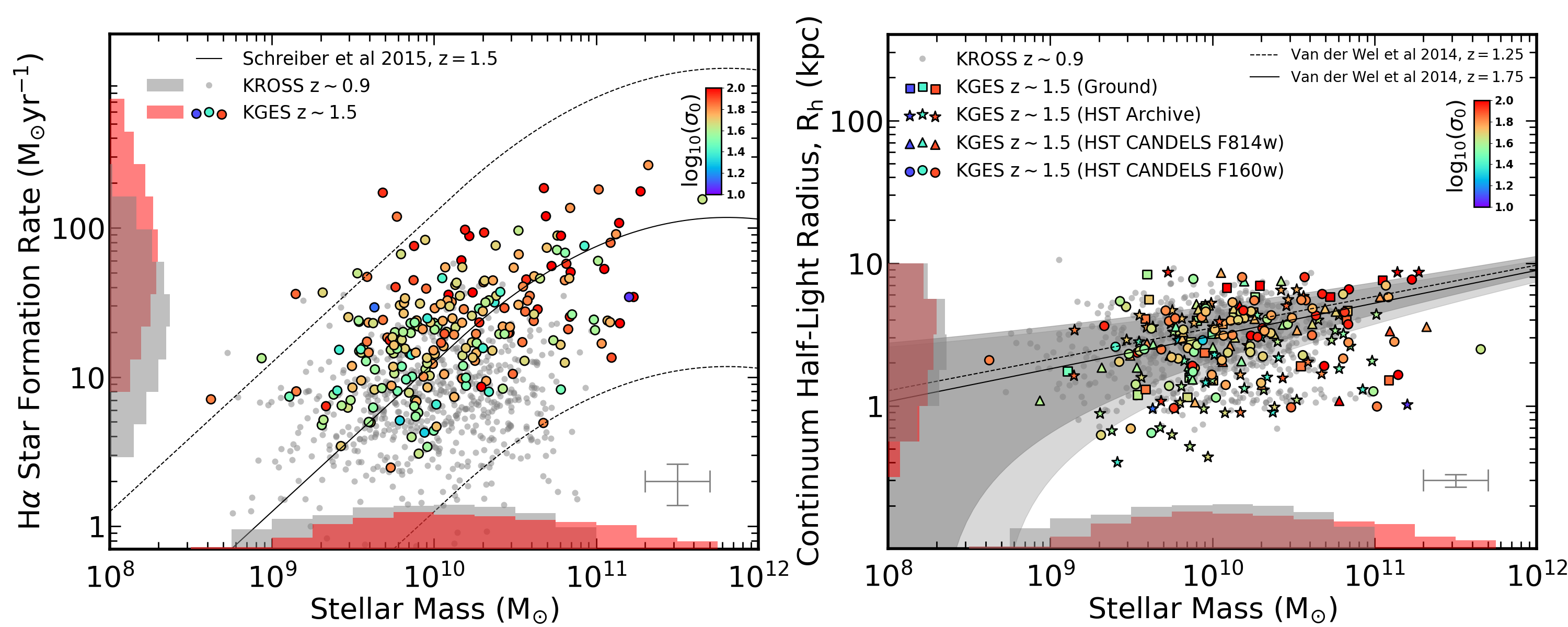}
	\caption{\emph{Left panel}: The extinction corrected H$\alpha$ star formation rate for the KGES sample as a function of stellar mass as derived from SED fitting using {\sc{magphys}} \citep{Cunha2008}. The KROSS $z$\,$\sim$\,0.9 sample is shown as grey points in the background. The \citet{Schreiber2015} $z$\,=\,1.5 star-formation rate stellar mass tracks, converted to a Chabrier IMF, are shown as well as factor 10 above and below the model track. 
	\emph{Right}: Stellar continuum half-light  radii, derived from {\sc{galfit}}, as a function of stellar mass. KROSS $z$\,$\sim$\,0.9 sample shown as grey points in the background. Ground (\emph{H}, \emph{K}) imaging (squares), non-CANDELS $HST$ imaging (stars), CANDELS $HST$ F814W imaging (triangles) and, CANDELS $HST$ F160W imaging (circles). The dashed and solid lines indicate the mass-size relation for star--forming galaxies at $z$\,=\,1.25 and $z$\,=\,1.75 respectively, as derived by \citet{VanderWel2014}, with the shaded region indicating the uncertainty on the relations. The median uncertainty on stellar mass, star formation rate and stellar continuum size are shown by grey bars in the lower right corner of each panel and the distribution of velocity dispersion within the sample is shown by the colour bar. In both panels we show histograms of each observable for both KROSS and KGES surveys. The figure indicates that the star-formation rates and stellar continuum sizes of the KGES galaxies are `typical' of star--forming galaxies at $z$\,$\sim$\,1.5.}
\label{Fig:MS}
\end{figure*}

The final data cube was generated by centering the individual frames according to the position of the point spread function (PSF) star, and then using an iterative 3-$\sigma$ clip mean average to reject pixels with cosmic ray contamination. For flux calibration, standard stars were observed each night either immediately before or after the science exposures. These were reduced in an identical manner to the science observations. Of the 288 observed galaxies, 243 were detected in H$\alpha$ emission and 235 have spatially resolved  H$\alpha$
emission with a median redshift of $\langle$\,$z$\,$\rangle$\,=\,1.48\,$\pm$\,0.01 ranging from $z$\,=\,1.22\,--\,1.76.

\section{Analysis and Results}\label{Sec:Analysis}

In the following sections we discuss galaxy integrated properties, (e.g. stellar mass  (M$_*$) and star-formation ($\dot{\rm M}_*$), stellar continuum half-light radius (R$_{\rm h}$) and S\'ersic index ($n$)). We then measure the galaxy dynamics and use the morphological properties, such as stellar continuum half-light radius, to extract and analyse the  galaxies' kinematic information.

\subsection{Stellar Masses and Star-Formation Rates}\label{Sec:MS}

Our targets were selected to lie in the ECDFS, UDS and COSMOS extragalactic fields prioritising the $HST$ CANDELS regions and therefore having a wealth of ancillary photometric data available. This allows us to construct spectral energy distributions (SEDs) for each galaxy spanning from the rest-frame $UV$ to mid-infrared with photometry from UDS \citep{Almaini2007}, COSMOS \citep{Muzzin2013} and ECDFS \citep{Giacconi2001}. 

To measure the galaxy integrated properties we derive the multi-wavelength photometry from  $UV$\,--\,8\,$\micron$ by cross correlating the galaxies in the KGES survey with the catalogs from the surveys listed above. The median the $U$, $I$ and $K$--band magnitude of the sample is $\langle$\,U$_{\rm AB}$\,$\rangle$\,=\,24.7\,$\pm$\,0.06, $\langle$\,I$_{\rm AB}$\,$\rangle$\,=\,23.7\,$\pm$\,0.04 and $\langle$\,K$_{\rm AB}$\,$\rangle$\,=\,22.2\,$\pm$\,0.06. We then use the {\sc{magphys}} \citep{Cunha2008,daCunha2015} code to fit spectral templates to the spectrum of each galaxy from which we derive stellar masses and dust attenuation factors (A$_{\rm v}$) (Dudzevi\v{c}i\={u}t\.{e} et. al. 2019). The full stellar mass range of our sample is $\log$(M$_{*}$[M$_{\odot}$])=8.9\,--\,11.7 with a median of $\log$(M$_{*}$[M$_{\odot}$])=10.0\,$\pm$\,0.1. We employ a homogeneous stellar mass uncertainty of $\pm$\,0.2 dex throughout this work  that conservatively accounts for the uncertainties in stellar mass values derived from SED fitting of high-redshift star--forming galaxies \citep{Mobasher2015}. We show the SEDs and {\sc{magphys}} fits for all galaxies in Appendix \ref{App:SEDs}.

The star formation rates of the galaxies in our sample are derived from the intensity of the summed H$\alpha$ emission--line fluxes in 2.4 arcsecond diameter apertures  in the KMOS observations. 
Following \cite{Wuyts2013}, we convert the dust attenuation  (A$_{\rm v}$), derived from {\sc{magphys}} SED fit for each galaxy, to a gas extinction correction factor. We assume a uniform uncertainty of $\pm$\,0.3 mag on the A$_{\rm v}$ of each galaxy to ensure the systematics in deriving dust attenuation factors from SED fitting are accounted for \citep{Muzzin2009}. We then derive extinction-corrected star formation rates for each galaxy following \cite{Calzetti2000}. The median H$\alpha$ star-formation rate of the galaxies in our sample is $\langle$\,SFR\,$\rangle$\,=\,17\,$\pm$\,2 M$_{\odot}$yr$^{-1}$ with a  16\,--\,84th percentile range of 3\,--\,44 M$_{\odot}$yr$^{-1}$. 


The H$\alpha$ star-formation rates and stellar masses for the KGES sample are shown in Figure \ref{Fig:MS}. For comparison we also show the KROSS $z$\,$\sim$\,0.9 sample (\citealt{Harrison2017}) as well as 0.1, 1 and 10$\times$ the `main-sequence' for $z$\,=\,1.5 star--forming galaxies derived in \citet{Schreiber2015}. The KGES sample is offset to higher H$\alpha$ star-formation rates compared with KROSS and reflects the increase in the cosmic star formation rate density at this epoch. We conclude that the galaxies in our sample at $z$\,$\sim$\,1.5 are representative of the star formation main--sequence at this redshift.

\subsection{Galaxy Morphology}\label{Sec:Phot}

To investigate the correlation between specific stellar angular momentum and morphology we need to quantify the morphology of the galaxies in our sample as well as derive their stellar continuum half-light radii. There are a variety of different approaches to classify a galaxy's morphology and in this section we derive both parametric and non-parametric classifications.

We first discuss the derivation and calibration of the S\'ersic index and stellar continuum half-light radius, using the {\sc{galfit}} software (\citealt{Galfit2011}), as well as analysis of the galaxy's axis ratios and inclinations. To quantify the morphologies non-parametrically, we also measure the Concentration, Asymmetry and Clumpiness (CAS; \citealt{Abraham1996,Conselice2014}) parameters for the galaxies in the KGES survey.

All of the galaxies in the sample were selected from the extragalactic deep fields, either UDS, COSMOS or ECDFS. Just over half the sample (162 galaxies) are part of the CANDELS survey, and so have have deep imaging in $VIJH$ wavelength bands, whilst 94 more have $HST$ archival imaging (mostly ACS $I$\,--\,band). For the remaining 32 galaxies we use ground based imaging to derive the morphological properties of the galaxies. 

The breakdown of broadband imaging available for the KGES sample, and the PSF half-light radius in each band, is given in Table \ref{Table:images}. At $z$\,=\,1.5, the observed near\,--\,infrared samples the rest frame $V$\,--\,band emission, red-ward of the 4000\,\r{A} break. To estimate the extent of the stellar light distribution, we use the longest wavelength $HST$ or ground-based image available.

\begin{table*}
\centering
\caption{The broadband imaging available for KGES galaxies that lie in the COSMOS, UDS and ECDFS fields. Survey, wavelength  band, number of galaxies, PSF FWHM and reference paper / programme ID are given. (CANDELS = The Cosmic Assembly Near-infrared Deep Extragalactic Legacy Survey. COSMOS = Cosmic Evolution Survey. UKIDDS = UKIRT Infrared Deep Sky Survey. TENIS = Taiwan ECDFS Near-Infrared Survey. UVISTA=Ultra Deep Survey near-infrared survey with VISTA telescope. $\dagger$\,=\,Ground based imaging.)}
\label{Table:images}
\begin{tabular}{l|l|c|c|l|}
\hline 
Survey & Band &  No. Gal. & PSF FWHM &Reference / Programme ID \\ 
\hline 
CANDELS & F435W, F606W, F814W & 112  & 0\farcs{22} &\citet{Koekemoer2011}, \citet{Grogin2011}\\ 
 & F105W, F125W, F160W &  & \\ 
CANDELS & F435W, F606W, F814W & 50 & 0\farcs{11}& \citet{Koekemoer2011}, \citet{Grogin2011}\\ 
$HST$ Archive & F140W &	 3 & 0\farcs{22} &  $HST$ ID:  13793 \\ 
$HST$ Archive & F125W &  3  & 0\farcs{22}& $HST$ ID:  15115 \\ 
COSMOS & F814W& 88  & 0\farcs{11}&  \citet{Koekemor2007}, \citet{Massey2010} \\
$^\dagger$COSMOS UVISTA DR3 & H &  3  & 0\farcs{76}& \citet{McCracken2012} \\
$^\dagger$UDS UKIDDS DR10& K 	&  22 & 0\farcs{77}& \citet{Lawrence2007}\\ 
$^\dagger$ECDFS TENIS & K &  7  & 0\farcs{91}& \citet{Bau2012} \\ 
\hline 
\end{tabular}
\end{table*}

\subsubsection{S\'ersic Index and Stellar Continuum Size}\label{Sec:Galfit}
We model the stellar light distributions of galaxies in the KGES sample, within 10\,$\times$\,10 arcsecond cutouts, using the {\sc{galfit}} software (\citealt{Galfit2011}) which fits single S\'ersic profiles of the functional form,
\begin{equation}\label{Sersic}
\rm I(r)=I_{e}\,exp\left[-b_{n}\left(\left(\frac{r}{\rm R_h}\right)^{1/{\it n}} - 1\right)\right],
\end{equation}
to the light profile of each galaxy. The S\'ersic index ($n$), is allowed to vary between $n$\,=\,0.2\,--\,8 and R$_{\rm h}$ defines the galaxy's stellar half-light radius. The S\'ersic models are convolved with the PSF of the broadband image, derived from stacking unsaturated stars in the frame. We show examples of the imaging, model and residuals for a sample of galaxies in Appendix \ref{App:Galfit}, as well the best quality image available for every KGES galaxy in Appendix \ref{App:SEDs}. For the galaxies with $HST$ CANDELS F160W coverage, we make a direct comparison of S\'ersic index (n), half-light radius (R$_{\rm h}$) and semi-major axis (PA) to \citet{Wel2012} who derived the structural properties of galaxies in the CANDELS survey up to $z$\,=\,3 also using {\sc{galfit}}. We find median ratios of $\langle$\,n$_{\rm {GF}}$/n$_{\rm {VW}}$\,$\rangle$\,=\,1.06$\,\pm$\,0.01, $\langle$\,R$_{\rm h_{GF}}$/R$_{\rm h_{VW}}$\,$\rangle$\,=\,1.00$\,\pm$\,0.01 and $\langle$\,PA$_{\rm {GF}}$/PA$_{\rm {VW}}$\,$\rangle$\,=\,1.00$\,\pm$\,0.01, where the subscript VW denotes \citet{Wel2012} measurements and GF denotes our measurement using {\sc{galfit}}.
This indicates that we can accurately recover the structural properties of $z$\,$\sim$\,1.5 galaxies using the {\sc{galfit}} software.

To ensure the measure of a galaxy's stellar continuum half-light radius is robust and unaffected by recent star--formation, we need measure the morphology of the galaxy in the longest wavelength band. To calibrate the structural properties of galaxies without $HST$ CANDELS F160W coverage, we use {\sc{galfit}} to fit S\'ersic profiles in every wavelength band that is available for each galaxy. We use the median ratios of half-light radius, S\'ersic index and semi-major axis in that band to the F160W wavelength band for galaxies with multi-wavelength imaging, to `correct' the structural properties to F160W measurements. At $z$\,=\,1.5 $HST$ F160W filters corresponds to $R$\,--\,band (640nm) whilst the $HST$ F814W samples the $U$\,--\,band (325nm) emission. To ensure the calibration of S\'ersic index is valid for galaxies of varying F814W-F160W colour (m$_{\rm F160W}$-m$_{\rm F814W}$), e.g. galaxies with more diverse stellar populations, we explore correlation between the S\'ersic index ratio $n_{\rm F160W}$\,/\,$n_{\rm F814W}$  and m$_{\rm F160W}$-m$_{\rm F814W}$ colour. 
We fit a linear function of the form,
\begin{equation}\label{Eqn:col}
\rm \frac{n_{\rm F160W}}{n_{\rm F814W}}=\alpha(m_{\rm F160W}-m_{\rm F814W})+\beta,
\end{equation} 
finding $\alpha$\,=\,$-$\,0.47 and $\beta$\,=\,0.64. On average, the ratio of S\'ersic index measured in F814W to F160W is $\langle$\,$n_{\rm F160W}$\,/\,$n_{\rm F814W}$\,$\rangle$\,=\,1.54\,$\pm$\,0.08 and this increases for galaxies with bluer colours.
We apply this variable calibration factor to the galaxies with $HST$ F814W imaging. The median S\'ersic index of KGES galaxies is $\langle$\,$n$\,$\rangle$\,=\,1.37\,$\pm$\,0.12, indicating their stellar light distributions are very similar to that of an exponential disc ($n$\,=\,1).

We also correct the stellar continuum half-light radii measured from F814W imaging, to equivalent F160W measurements, following a similar procedure and deriving a fixed correction factor of $\langle$\,R$_{\rm h,F160W}$\,/\,R$_{\rm h,F814W}$\,$\rangle$\,=\,0.90\,$\pm$\,0.02. This indicates that, on average, the stellar continuum sizes measured from F814W band imaging are 10 per cent larger than that measured from F160W band imaging. 
We derive a median intrinsic R$_{\rm h}$ of the galaxies in our sample to be $\langle$\,R$_{\rm h}$\,$\rangle$\,=\,0\farcs{31}$\, \pm$\,0\farcs{02} (2.60\,$\pm$\,0.15\,kpc at  $z$\,=1.5). In Figure \ref{Fig:MS} we show the distribution of half-light radius (R$_{\rm h}$), derived from a variety of imaging (Table \ref{Table:images}) as a function of stellar mass for all 288 KGES galaxies. We show tracks of the stellar mass - stellar continuum size relation from \citet{VanderWel2014} for star--forming galaxies at $z$\,=\,1.25 and $z$\,=\,1.75  with the shaded region indicating the uncertainty on the relations. The main--sequence galaxy population, in the redshift range $z$\,=\,1.25\,--\,1.75, with a median stellar mass of $\log$(M$_{*}$[M$_{\odot}$])=10.25, has stellar continuum size 18\,--\,64th percentile range of $\langle$\,R$_{\rm h}$\,$\rangle$\,=\,1.32\,--\,5.5\,kpc \citep{VanderWel2014}. The median size of the KGES galaxies lies within this range and from Figure \ref{Fig:MS} we can see that the galaxies in the KGES survey have stellar continuum sizes that are typical of the star--forming population at  $z$\,=1.5.

To place the KGES sample in context of other high--redshift integral field studies of star--forming galaxies, we also show the stellar continuum size distribution of the KROSS survey as a function of stellar mass in Figure \ref{Fig:MS}. The distribution of sizes in the two surveys is very similar with KROSS having a slightly larger a median size of $\langle$\,R$_{\rm h}$\,$\rangle$\,=\,0\farcs{36}$\, \pm$\,0\farcs{01} (2.80\,$\pm$\,0.07\,kpc at $z$\,=\,0.9). 

\begin{figure*}
	\centering
	\includegraphics[width=1\linewidth,trim={1cm 2cm 1cm 0}]{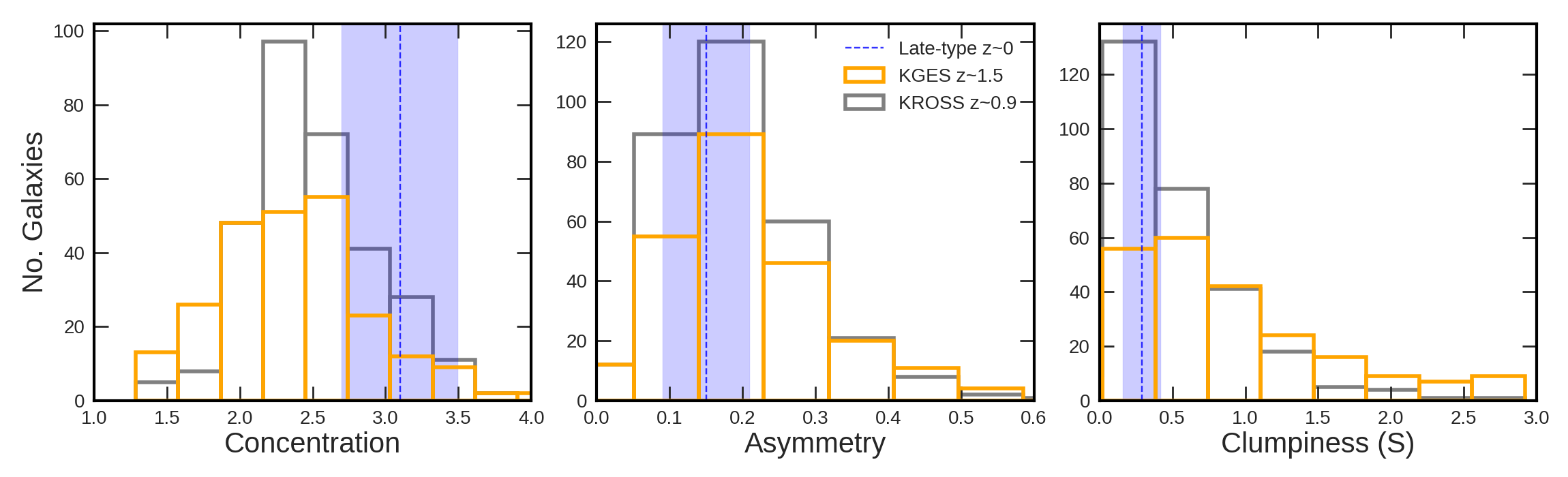}
	\caption{Histograms of the Concentration, Asymmetry and Clumpiness of the KGES $z$\,$\sim$\,1.5 galaxies (orange) measured from $HST$ F814W imaging.  We also show the distribution the KROSS $z$\,$\sim$\,0.9 survey \citep{Harrison2017} with $HST$ F814W imaging (grey) as well as the median values and scatter (blue line and shaded region) for a sample of late--type $z$\,=\,0 galaxies from \citet{Conselice2003} who used $R$\,--\,band imaging. The KGES galaxies are comparable in concentration and asymmetry to KROSS, whilst being clumpier on average. The $z$\,=\,0 sample is more concentrated and less clumpy than KGES whilst having similar asymmetry. }
\label{Fig:CAS_Hist}
\end{figure*}

\subsubsection{Inclination and Axis Ratios}

In Section \ref{Sec:Kin} we will measure the rotational velocities of the galaxies in the sample. To correct the dynamics for line-of-sight inclination effects we derive the inclination for each galaxy in the sample. For galaxies that are disc-like, the inclination angle can be calculated using,
\begin{equation}\label{Eqn:Inc}
\rm{cos}^2(\theta_{\rm inc })=\frac{\left({\rm b }/{\rm a }\right)^2-\rm q_0^2}{1-\rm q_0^2},
\end{equation} 

\noindent where $\theta_{\rm inc }$\,=\,0 represents a face-on galaxy (e.g.\,\citealt{TF1977}). The value of q$_0$, which represents the edge on axis ratio, depends on the galaxy type, but is typically in the range  q$_0$\,=\,0.13\,--\,0.20 for rotationally supported galaxies at $z$\,$\sim$\,0 (e.g. \citealt{Weijmans2016}). We adopt q$_0$\,=\,0.2 as this is appropriate for a thick disc (e.g. \citealt{Guthrie1992,Law2012,Weijmans2014}) and to be consistent with other high-redshift integral field surveys (e.g. KROSS, \citealt{Harrison2017}; KMOS3D, \citealt{Wisnioski2015}). The medium axis-ratio of KGES galaxies, derived from the {\sc{galfit}} modelling, is $\langle$\,b/a$\,\rangle$\,=\,0.60\,$\pm$\,0.02 which equates to a medium inclination of $\langle\, \theta_{\rm inc}\,\rangle$\,=\,55$^{\circ}$\,$\pm$\,2$^{\circ}$. This corresponds to a medium line-of-sight velocity correction of $\sim$\,30 percent. To measure the reliability of the axis ratio measurements from {\sc{galfit}} for the KGES galaxies, we generate 1000 mock galaxies with a distribution of half--light radii, S\'ersic index, $K$--band magnitude and axis ratios that reflects the KGES sample. We use \textsc{galfit} to measure the intrinsic axis ratio of the model galaxies and derive a median ratio of $\langle$\,ba$_{\rm  \,int}$\,/\,ba$_{\rm\, GALFIT}$\,$\rangle$\,=\,1.00\,$\pm$\,0.01 with a scatter of 0.40. We note however that {\sc{galfit}} performs poorly for very faint small galaxies that have low signal to noise.
The median axis ratio is in agreement with the results of \citet{Law2012a} who use the rest-frame $HST$ optical images for $z$\,$\approx$\,1.5\,--\,3.6 star--forming galaxies and find a peak axis ratio of (b/a)$\sim$0.6.

\subsubsection{Concentration, Asymmetry and Clumpiness (CAS)} \label{Sec:CAS}

In Section \ref{Morph_j} we will correlate the dynamics of the galaxies with their morphologies, so to provide a non--parametric model independent measurement of a galaxies rest-frame optical morphology, we next derive the Concentration, Asymmetry and Clumpiness (CAS; \citealt{Abraham1996,Conselice2003,Conselice2014}) of the continuum stellar light distribution of the galaxies in our sample. 
As shown by \citet{Conselice2003}, due to the their non-parametric nature, the CAS parameters of star--forming galaxies can be reliably measured out to high redshift and they capture the major features of the stellar structure in a galaxy without assuming an underling form, e.g. S\'ersic fitting in the case of {\sc{galfit}}. We note due to the complex, non-linear, nature of converting non-parametric measures of a galaxies morphology between different wavelength bands, we do not measure the CAS parameters for galaxies without $HST$ imaging. For galaxies with $HST$ imaging, we derive the CAS parameters in F814W $I$\,--\,band imaging as this maximises the sample size and allows an accurate comparison to the KROSS survey which predominately has $HST$ F814W $I$\,--\,band imaging.

The Concentration (C) of a galaxy is a measure of how much light is in the central regions of the galaxy compared to the outskirts and is calculated from,
\begin{equation}
\rm C=5 \times \log_{10} \left( \frac{r_{outer}}{r_{inner}}\right),
\end{equation}
\noindent where r$_{\rm outer}$ is the radius which contains  80 per cent of the light within an aperture of semi-major axis 3R$_{\rm h}$, r$_{\rm inner}$ is the radius which contains 20 per cent of the light within the same aperture. A higher value of concentration indicates a larger fraction of the galaxies light originates from the central regions. 
The median concentration for our sample is $\langle$\,C\,$\rangle$\,=\,2.36\,$\pm$\,0.34.
For comparison we also measured the concentration of galaxies in the KROSS $z$\,=\,0.9 sample with $HST$ imaging (178 galaxies), finding  $\langle$\,C\,$\rangle$\,=\,2.4\,$\pm$\,0.27 which implies, on average the stellar light profiles of $z$\,=\,0.9 star--forming galaxies are more concentrated than $z$\,=\,1.5 galaxies. \citet{Conselice2003} identified that in a sample of 250 $z$\,$\sim$\,0 galaxies, late-type discs have a median concentration of $\langle$\,C\,$\rangle$\,=\,3.1\,$\pm$\,0.4, whilst local early type galaxies have much higher concentration of $\langle$\,C\,$\rangle$\,=\,3.9\,$\pm$\,0.5. Local irregular galaxies were established to have a $\langle$\,C\,$\rangle$\,=\,2.9\,$\pm$\,0.3 indicating high--redshift galaxies have stellar light distributions with concentrations similar to local irregular galaxies.

The Asymmetry (A) of a galaxy  reflects the fraction of light originating from non-symmetric components, where a perfectly symmetric galaxy would have A\,=\,0 and a maximally asymmetric galaxy would have A\,=\,1. The Asymmetry estimator of a galaxy is defined as,
\begin{equation}
\rm A=min \left( \frac{\sum |I_0 -I_{180}|}{\sum |I_0|}\right) -  min \left( \frac{\sum |B_0 -B_{180}|}{\sum |I_0|}\right),
\end{equation}

\noindent where I$_{0}$ represents the original galaxy image and I$_{180}$ is the image rotated by 180$^\circ$ about its centre. B$_{0}$ and B$_{180}$ represent a region of sky of equal size nearby to the galaxy \citep{Conselice2014}. 
The true Asymmetry of the galaxy is measured by minimising over the centre of symmetry and is calculated within an ellipse of semi--major axis 3R$_{\rm h}$, where R$_{\rm h}$ is convolved with the PSF of the image, with an axis ratio and position angle matching that derived from S\'ersic fitting in Section \ref{Sec:Galfit}. 

Since the Asymmetry is a function of signal to noise \citep{Conselice2003}, we assess the reliability of Asymmetry measurements by creating 100 mock galaxies with S\'ersic index $n$\,=\,0.5\,--\,2, R$_{\rm h}$\,=\,0\farcs{1}\,--\,1\farcs{0} and a signal to noise distribution similar to our data. The  Asymmetry in each galaxy is calculated first within  an ellipse of semi--major axis 3R$_{\rm h}$ (A$_{\rm Mask}$) and compared to the true Asymmetry of each galaxy (A$_{\rm True}$), derived from the full extent of the galaxy with infinite signal to noise. We then compare  A$_{\rm True}$ to the Asymmetry within  an ellipse of semi--major axis 3R$_{\rm h}$ for galaxies that have signal to noise of 10 (A$_{10}$).
We find a median ratio of $\langle$\,A$_{\rm True}$\,/\,A$_{\rm Mask}$\,$\rangle$\,=\,1.01\,$\pm$\,0.03 whilst $\langle$\,A$_{\rm True}$\,/\,A$_{\rm 10}$\,$\rangle$\,=\,1.05\,$\pm$\,0.01. This indicates that on average the Asymmetry of the galaxies, although slightly underestimated, are accurate to a few per cent when calculated within an ellipse of semi--major axis 3R$_{\rm h}$, even in our lowest signal to noise sources.

For the KGES galaxies we derive a median Asymmetry of $\langle$\,A\,$\rangle$\,=\,0.19\,$\pm$\,0.05 with a range from A\,=\,0.01\,--\,0.85. In a study of $z$\,$\sim$\,0 galaxies by \citet{Conselice2003}, late--type galaxies have  $\langle$\,A\,$\rangle$\,=\,0.15\,$\pm$\,0.06, whilst early--types have $\langle$\,A\,$\rangle$\,=\,0.07\,$\pm$\,0.04 and irregular galaxies have $\langle$\,A\,$\rangle$\,=\,0.17 \,$\pm$\,0.10. The galaxies in the KGES survey have asymmetries equivalent to those of local late--type and irregular galaxies. In Section \ref{Sec:AM_corr} we will also compare the dynamics and morphology of the KROSS sample to the KGES galaxies. We therefore derive the Asymmetry of the KROSS galaxies, finding  $\langle$\,A\,$\rangle$\,=\,0.16 \,$\pm$\,0.06.

We can parameterise the fraction of light originating from clumpy distributions in a galaxy using the Clumpiness parameter. S, which is defined as,
\begin{equation}
\rm S=10 \times \left[\left( \frac{\sum (I_{x,y} -I_{x,y}^{\sigma})}{\sum I_{x,y}}\right) - \left( \frac{\sum B_{x,y} -B_{x,y}^{\sigma}}{\sum I_{x,y}}\right)\right],
\end{equation}
\noindent where  $\rm I_{x,y}$ is the original image and  $\rm I_{x,y}^{\sigma}$ is a smoothed image. The degree of smoothing, as defined by \citet{Conselice2003}, is relative to the size of the galaxy and is quantified by $\sigma$\,=\,0.2$\,\times$\,3R$_{\rm h}$, where $\sigma$ is the standard deviation of the Gaussian kernel. The residual map generated  from subtracting the smoothed image from the original, contains only high frequency structures in the galaxy. The central region of the galaxy is masked out in this process as it is often unresolved.

The same method is applied to an arbitrary region of background away from the galaxy ($\rm B_{x,y}$, $\rm B_{x,y}^{\sigma}$) to remove the inherent Clumpiness of the noise in the image. We derive the Clumpiness for the galaxies in the KGES sample finding a median Clumpiness of $\langle$\,S\,$\rangle$\,=\,0.37\,$\pm$\,0.14 with a range from S\,=\,0.01\,--\,5.3. In comparison to the local Universe, \citet{Conselice2003} identified that $z$\,$\sim$\,0 late--type galaxies have $\langle$\,S\,$\rangle$\,=\,0.29\,$\pm$\,0.13, early--type galaxies have $\langle$\,S\,$\rangle$\,=\,0.08\,$\pm$\,0.08 and irregular galaxies have $\langle$\,S\,$\rangle$\,=\,0.40\,$\pm$\,0.20. The Clumpiness distribution of KGES galaxies aligns with that of late--type local disc galaxies, although we note that a larger will reduce the clumpiness measured in a galaxy. As a comparison sample we also derive the Clumpiness for the galaxies in the KROSS sample, finding a median value of $\langle$\,S\,$\rangle$\,=\,0.37\,$\pm$\,0.10.   

\citet{Law2012a} established that a typical main--sequence star--forming galaxy in the redshift range $z$\,=\,1.5\,--\,3.6 is well described by a S\'ersic profile of index $n$\,$\sim$\,1, Concentration index C$\sim$\,3 and Asymmetry index A$\sim$\,0.25. The galaxies in the KGES sample have S\'ersic and CAS parameters that align with typical star--forming galaxies at $z$\,=\,1.5. We show the distribution of Concentration, Asymmetry and Clumpiness of the KGES $z$\,$\sim$\,1.5 galaxies in comparison to the KROSS $z$\,$\sim$\,0.9 survey as well as the median values and scatter  for a sample of late--type $z$\,=\,0 galaxies from \citet{Conselice2003} in Figure \ref{Fig:CAS_Hist}.

\begin{figure*}
	\centering
	\includegraphics[width=1\linewidth,trim={0 2cm 0 0}]{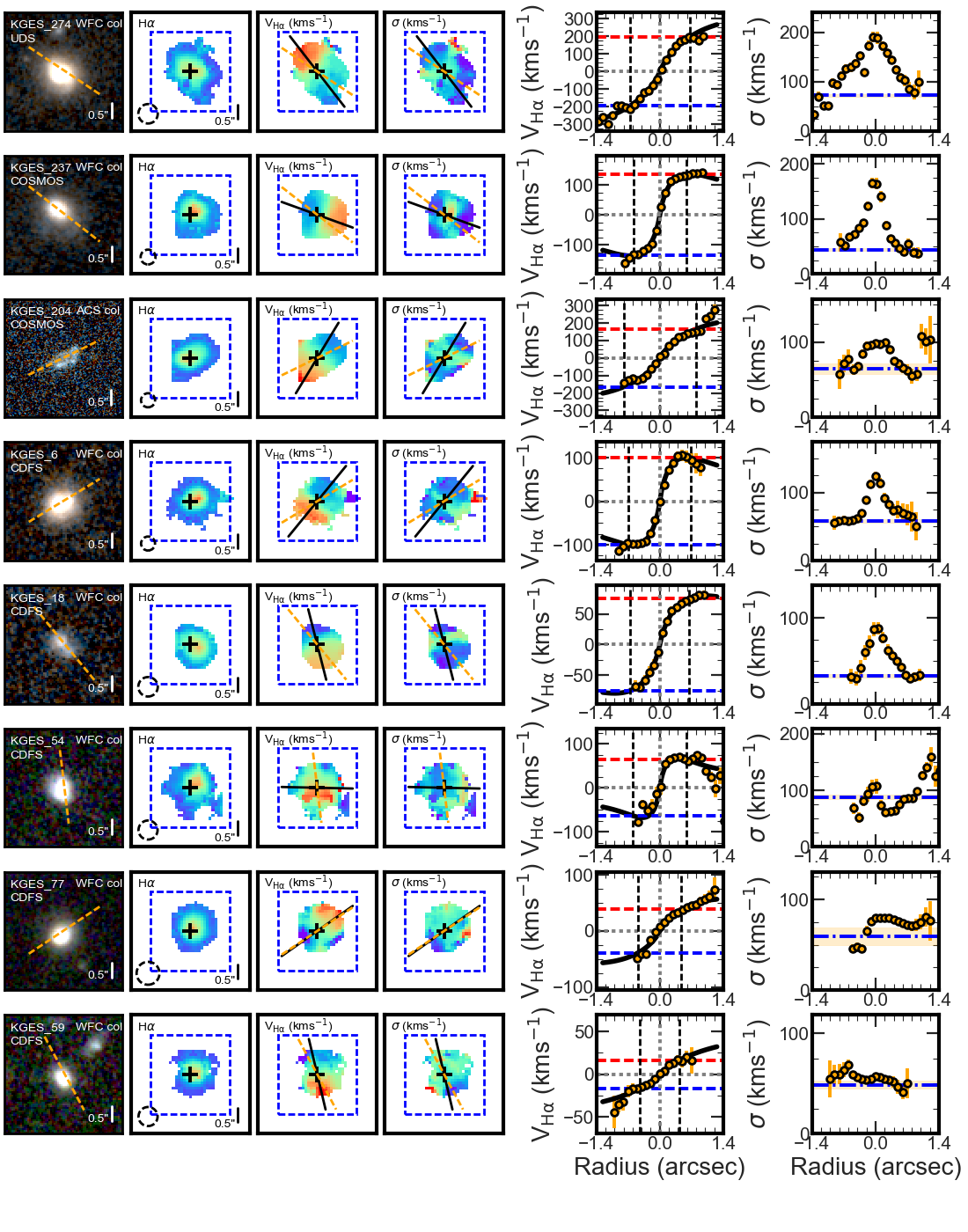}
	\caption{Example of spatially resolved galaxies in the KGES sample from each quartile of specific stellar angular momentum. From left to right: broad-band imaging of the galaxy (left), with semi-major axis (PAim; orange dashed line), H$\alpha$ intensity map, velocity map, and velocity dispersion map, derived from the
emission-line fitting with data cube field of view (blue dashed square). Kinematic position angle (PAvel; black solid line) and PAim (orange dashed line) axes are plotted on the rotation and dispersion velocity maps. Rotation curve  and dispersion profile extracted about the kinematic position axis (right). The rotation curve shows lines of 2R$_{\rm h}$ derived from S\'ersic fitting, as well as V(2R$_{\rm h}$) (red and blue dashed lines) extracted from the rotation curve fit (black curve). The dispersion profile shows the extracted $\sigma_{\rm int}$ (blue dashed line) and 1$\sigma$ region (yellow shaded region).}
\label{Fig:kin_pg0}
\end{figure*}

\begin{figure*}
	\centering
	\includegraphics[width=0.9\linewidth]{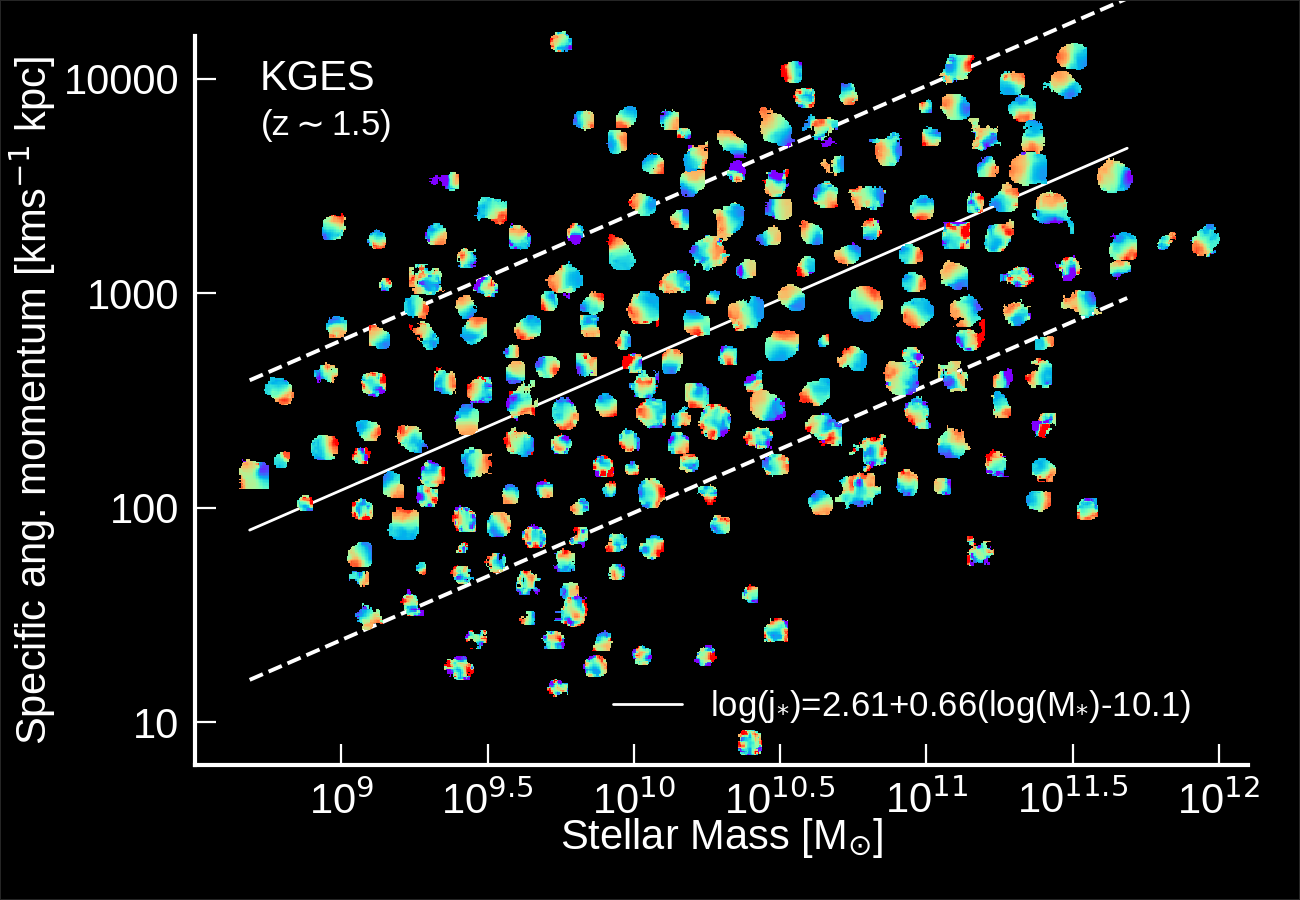}
	\caption{The H$\alpha$ rotational velocity maps of the KGES galaxies with H$\alpha$ signal to noise greater than 5, displayed in the specific stellar angular momentum stellar mass plane, offset to minimise overlap.	The white solid line is a fit to the KGES data of the form $\log_{10}(j_*)$\,=\,$\alpha$\,+$\,\beta\,(\log_{10}(\rm M_*/M_{\odot})-10.10)$, with the slope fixed to $\beta$\,=\,0.66 and a derived intercept of $\alpha$\,=\,2.61. White dashed lines are a factor of 10 above and below the fit. Low stellar mass, low angular momentum galaxies have smaller stellar continuum sizes and thus have a smaller extent of nebula emission compared to galaxies of higher stellar mass and higher angular momentum.}
\label{Fig:MS_poster_JM}
\end{figure*}

\subsection{Kinematics} \label{Sec:Kin}

We next turn our attention to the kinematics of the KGES sample. A full description of the emission--line fitting procedure and extraction of kinematic properties is given in Tiley et. al. (in prep.). Here we give a brief overview of the emission--line fitting procedure and then we discuss the rotational velocity and velocity dispersion measurements that enable us to quantify more derived properties of the KGES galaxies.

\subsubsection{Emission--Line Fitting}

Briefly, we fit a triple Gaussian profile to the continuum subtracted H$\alpha$ ($\lambda$6562\,\AA)  and [{N\sc{ii}}] ($\lambda$6548\,\AA, $\lambda$6583\,\AA) emission line profiles in all 288 KGES galaxies, with the redshift, emission--line width and emission--line amplitude as free parameters. The three emission lines share a common width and their relative positions are fixed according to \citet{Osterbrock2006}. The instrumental broadening of the OH sky lines by KMOS is used to correct for instrumental broadening 

For each galaxy, we fit  the emission--line profiles in the integral field observation  using an adaptive binning technique. Starting in apertures of 0.3\,$\times$\,0.3 arcsecond (comparable to half the FWHM of the seeing), we impose a H$\alpha$ signal to noise threshold of S/N\,$\geq$\,5 on the integrated S/N of the emission line. If this S/N is not achieved, we fit to the spectrum over a larger area until either the S/N threshold is achieved or the binning limit of 0.7\,$\times$\,0.7 arcsecond (comparable to the FWHM of the seeing) is reached. In Figure \ref{Fig:kin_pg0} we show examples of the spatially resolved H$\alpha$ intensity, velocity, and velocity dispersion maps for a number of KGES galaxies. The H$\alpha$ velocity for all KGES galaxies in shown in Appendix \ref{App:SEDs}. The galaxies in our sample have predominantly rotationally supported gas kinematics, with $\langle$\,V$_{\rm 2R_h}$/$\sigma_0$\,$\rangle$\,=\,1.93\,$\pm$\,0.21 where 68 per cent of KGES galaxies have $v/\sigma>$1, within which V$_{\rm 2R_h}$ is the rotation velocity of the galaxy and $\sigma_0$ is the intrinsic velocity dispersion, as defined in Section \ref{Sec:V_2rh} $\&$ \ref{Sec:sigma}.
To quantify the misalignment between the kinematic and morphological we define the misalignment parameter $\Psi$ as, 
\begin{equation}
 \rm sin\Psi=|sin(PA_{\rm morph}-PA_{\rm kin})|
\end{equation}
where $\Psi$ is defined between 0$^{\circ}$ and 90$^{\circ}$ \citep{Wisnioski2015}. For the KGES sample $\langle$\,$\Psi$\,$\rangle$\,=\,18.65$^{\circ}$\,$\pm$ \,1.98$^{\circ}$ with 66 per cent of KGES galaxies passing the galaxy disc criteria of $\Psi < 30^{\circ}$. This fraction increases to 78 per cent with $\Psi < 40^{\circ}$. This indicates that the KGES galaxies have well defined velocity gradients, that reflect the stellar morphology shown in the first panel of Figure \ref{Fig:kin_pg0}. This indicates that most of the high--redshift galaxies in the KGES sample are predominantly rotation dominated galaxies with defined rotation axes. The distribution of H$\alpha$ velocity maps for the full sample in the specific stellar angular momentum stellar mass plane is shown in Figure \ref{Fig:MS_poster_JM}. We note however, that some `disc' galaxies in seeing-limited observations have been identified as mergers in higher resolution adaptive optics observations (e.g.\,\citealt{Rodrigues2017,Sweet2019}; Espejo et al. in prep.).

\subsubsection{Rotation Velocities}\label{Sec:V_2rh}

To measure the correlation between the dynamics of the galaxies in our sample and their rest frame optical morphologies, we need to parameterise their kinematics. We quantify the dynamics by measuring the asymptotic rotational velocity of each galaxy derived from the spatially resolved H$\alpha$ velocity maps.

The rotation curve of a galaxy is defined as the velocity profile extracted about the galaxy's kinematic position angle. For each galaxy, we measure the kinematic position angle by rotating the velocity map in one degree increments about the galaxy's continuum centre (defined from $HST$). For each step we calculate the velocity gradient along a horizontal `slit' of width equal to half the FWHM of the PSF of the seeing. We define the kinematic position angle as the average of the angle with maximum velocity gradient and the angle of minimum velocity gradient plus 90 degrees. We extract the velocity profile at the kinematic position angle, with the velocity and uncertainty taken as the weighted mean and standard error along pixels perpendicular to the `slit'. 

We choose this  method to derive the rotation profiles of the galaxies in the KGES sample as opposed to forward modelling approaches (e.g. \citealt{DiTeodoro2016}) since this reduces the number of assumptions about the galaxy's dynamical state. We note, however in doing so the extracted rotation curves are effected by beam smearing but by following the procedures of \citet{Johnson2018} these effects can be reduced to less than the 10 per cent level.

To minimize the scatter in the velocity profiles and to allow for the possibility of rising, flat or declining rotation curves, we fit each galaxy's rotation curve with a parametric model. We choose an exponential light profile (see \citealt{Freeman1970a}) since the kinematics, as shown in Figure \ref{Fig:kin_pg0}, indicate the majority of the galaxies are rotationally supported with large scale ordered rotation. The dynmaical model is parameterised as follows,
\begin{equation}\label{vcirc}
\rm v(r)^2\,=\,\frac{r^2\pi G \mu_o}{r_D}(I_o({\it x})K_o({\it x})-I_1({\it x})K_1({\it x}))
\end{equation}
\noindent where G is the gravitational constant, $\mu_o$ is the peak mass surface density, $r_{\rm D}$ is the disc scale radius and I$_{\rm n}(x)$K$_{\rm n}(x)$ are Bessel functions evaluated at $x$\,=\,0.5r/$r_{\rm D}$. The rotation velocities and best fit dynamical models are shown in Figure \ref{Fig:kin_pg0} for a subsample of KGES galaxies. We do not interpret the model parameters, nor extrapolate the model to large radii, but rather use the model to trace the observed rotational velocity profiles and account for the effect of noise in the outer regions.

Next we measure the rotational velocity of each galaxy by extracting the velocity from the galaxy's rotation curve at 2R$_{\rm h}$ (= 3.4R$_{\rm d}$ for an exponential disc where R$_{\rm d}$ is the light profile scale radius; e.g. \citealt{Miller2011}). As shown by \citet{Romanowsky2012}, the velocity at 2R$_{\rm h}$ provides a reliable estimate of a galaxy's rotation velocity irrespective of its morphology. At 2R$_{\rm h}$, the velocity profile of an exponential disc, with a nominal dark matter fraction, begins to flatten and the effects of beam smearing are minimized. It is also crucial for capturing the majority of a galaxy's angular momentum (e.g. \citealt{Obreschkow2015}), as we demonstrate in Section \ref{Sec:AM} for the KGES galaxies and allows comparison to other spatially resolved studies of star--forming galaxies 
\citep[e.g. KMOS$^{\rm 3D}$, KROSS,][]{Wisnioski2015,Harrison2017}

The extracted velocity, from the dynamical model, is inclination and beam smear corrected following the procedures described in \citet{Johnson2018} with a median correction factor of $\langle$\,V$_{\rm obs}$/V$_{\rm int}$\,$\rangle$\,=\,1.05\,$\pm$\,0.01. The median intrinsic rotation velocity of the KGES galaxies is $\langle$\,V$_{\rm 2R_h}$\,$\rangle$\,=\,102\,$\pm$\,8\,km\,s$^{-1}$, with a 16-84th percentile range of 27\,--\,191\,km\,s$^{-1}$. 

For 50 of the galaxies in the KGES sample, the low S/N of the H$\alpha$ emission, means we do not spatially resolve the galaxy out to 2R$_{\rm h}$. In these galaxies, we extrapolate the dynamical model beyond the last data point to measure the rotation velocity at 2R$_{\rm h}$.
To understand whether this affects the derived rotation velocity we measure the ratio of the radius of the last data point on the rotation curve to  2R$_{\rm  h}$ and the ratio of the velocity of the last data point to the velocity extracted at 2R$_{\rm  h}$. For galaxies we do resolve, we identify that $\langle$\,R$_{\rm last}$/2R$_{\rm h}$\,$\rangle$\,=\,1.6\,$\pm$\,0.08 and $\langle$\,V$_{\rm last}$/V$_{\rm 2R_h}$\,$\rangle$\,=\,1.01\,$\pm$\,0.03, whilst for the 50 galaxies we do not resolve out to 2R$_{\rm  h}$, $\langle$\,R$_{\rm last}$/2R$_{\rm h}$\,$\rangle$\,=\,0.84\,$\pm$\,0.04 and $\langle$\,V$_{\rm last}$/V$_{\rm 2R_h}$\,$\rangle$\,=\,0.97\,$\pm$\,0.02. This indicates that on average when the H$\alpha$ rotation curve does not extend out to 2R$_{\rm h}$, a 15 per cent extrapolation is required and the extracted velocity at 2R$_{\rm h}$ is slightly less than that at R$_{\rm last}$.

To put the dynamics of the galaxies in the KGES sample in the context of other high-redshift star--forming galaxy surveys, we make a comparison to the KROSS sample of $\sim$600 star--forming galaxies at $z$\,$\sim$\,0.9. \citet{Harrison2017} extracts the rotation velocity of the KROSS galaxies at 2R$_{\rm  h}$ and applying the beam smearing corrections derived in \citet{Johnson2018}. The KROSS sample has a median intrinsic rotational velocity of $\langle$\,V$_{\rm int}$\,$\rangle$\,=\,117\,$\pm$\,4\,km\,s$^{-1}$ with a 16-84th percentile range of 46\,--\,205\,km\,s$^{-1}$. In the KROSS sample, galaxies have higher rotation velocities than the KGES galaxies at $z$\,$\sim$\,1.5.

The distribution of stellar mass in both the KROSS and KGES surveys is very similar with both samples having a median stellar mass of $\log$(M$_{*}$[M$_ {\odot}$])=10.0\,$\pm$\,0.2. The origin of the  evolution in rotation velocities may be driven by the biases in the selection function of the two surveys or by an evolution in pressure support within the galaxies (e.g. \citealt{Tiley2019}, \citealt{Ubler2019}). Establishing the exact cause is beyond the scope of this paper, but will be discussed in Tiley et al. (in prep.).

\begin{figure}
	\centering
	\includegraphics[width=1\linewidth]{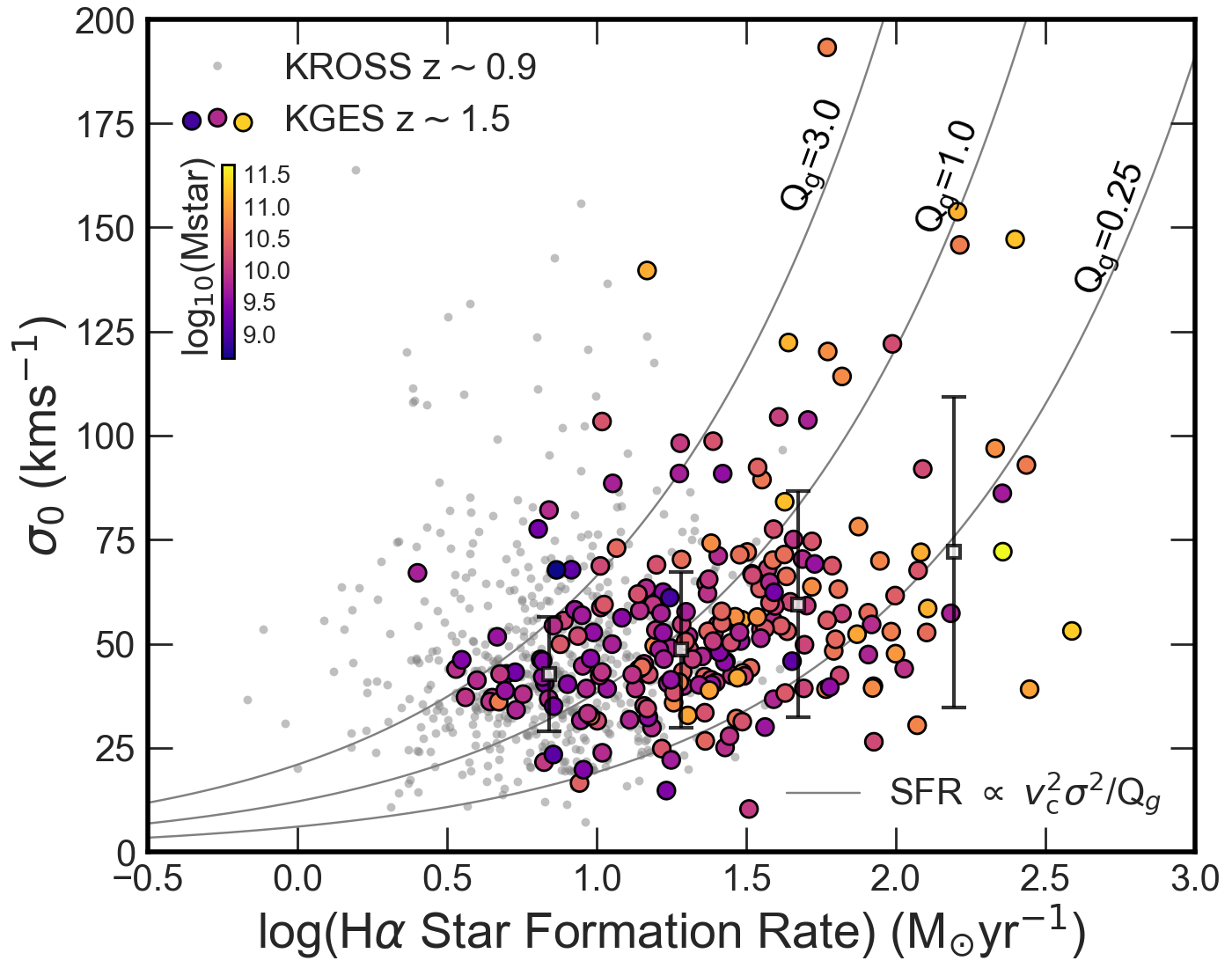}
	\caption{Velocity dispersion ($\sigma_0$) as a function of the H$\alpha$ star formation rate for KGES (coloured points) and KROSS (grey points) galaxies. KGES galaxies are coloured by their stellar mass (M$_*$) with the median and standard deviation of velocity dispersion in bins of H$\alpha$ star formation rate shown by the square points.  Galaxies of a higher star formation rate have higher stellar mass (Figure \ref{Fig:MS}). We show the feedback driven turbulence model from \citet{Krumholz2016} for the relation between star formation rate and velocity dispersion, parameterised as SFR$\,\propto$\,$v_{\rm c}^2\sigma^2/Q_{\rm g}$, for different Toomre Q$_{\rm g}$ values, evaluated at the median rotational velocity of the KGES sample, $\langle$\,V$_{\rm 2R_h}$\,$\rangle$\,=\,102\,$\pm$\,8\,km\,s$^{-1}$. The KGES galaxies occupy similar $\sigma_o$\,--\,SFR parameter space as galaxies with Q$_{\rm g}$\,=\,0.25\,--\,3.0 }
\label{Fig:sigma_sSFR}
\end{figure}

\subsubsection{Velocity dispersion}\label{Sec:sigma}

To analyse the connection between a galaxy's rest--frame optical morphology, dynamics and the balance between rotational and pressure support, we need to measure the intrinsic velocity dispersion (disc thickness) within each galaxy. We assume that a galaxy's intrinsic dispersion profile is flat and that the velocity dispersion is a good proxy for the turbulence (non-circular motions) within a galaxy.


We attempt to measure the dispersion profile of each galaxy out to 1.3R$_{\rm h}$. We choose 1.3R$_{\rm h}$ as opposed to 2R$_{\rm h}$, as more galaxies have kinematic information at 1.3R$_{\rm h}$ and we identify that the derived velocity dispersion is very similar with $\langle$\,$\sigma_{\rm 1.3R_h}$\,/\,$\sigma_{\rm 2R_h}$\,$\rangle$\,=\,1.00\,$\pm$\,0.07. If the spatially resolved kinematics of the galaxy do not extend out to  1.3R$_{\rm h}$, we measure the median dispersion from the velocity dispersion map of the galaxy, examples of which are shown in Figure \ref{Fig:kin_pg0}. The extracted values are then corrected  for beam smearing following the methods described in \citet{Johnson2018}, which use model-based corrections, to derive an intrinsic velocity dispersion for each galaxy.

For the sample of 235 resolved galaxies the median line-of-sight velocity dispersion is $\langle$\,$\sigma_{0}$\,$\rangle $\,=\,52\,$\pm$\,2\,km\,s$ ^{-1}$, with a 16-84th percentile range of 37\,--\,72\,km\,s$^{-1}$. In comparison, the KROSS sample of galaxies at $z$\,$\sim$\,0.9 has a median velocity dispersion of $\langle$\,$\sigma_{\rm 0}$\,$\rangle$\,=\,44\,$\pm$\,1\,km\,s$^{-1}$. \citet{Ubler2019} established that 
star--forming galaxies at $z$\,=\,2.3 have a ionized gas velocity dispersion of $\langle$\,$\sigma_{\rm 0}$\,$\rangle$\,=\,45\,km\,s$^{-1}$, whilst for galaxies at $z$\,=\,0.6, $\langle$\,$\sigma_{\rm 0}$\,$\rangle$\,=\,30\,km\,s$^{-1}$. This indicates that main sequence star--forming galaxies at $z$\,$\sim$\,1.5 have 20 per cent higher levels of turbulence compared to $z$\,$\sim$\,0.9 main sequence galaxies whilst having comparable levels of dispersion to higher redshift galaxies. This is in agreement with the findings of previous high redshift integral field studies (e.g. \citealt{Wisnioski2015,Johnson2018,Ubler2019}, Tiley et. al. in prep.).

In Figure \ref{Fig:sigma_sSFR} we show the velocity dispersions of both the KGES and KROSS galaxies as a function of their H$\alpha$  star formation rate, with the KGES galaxies coloured by their stellar mass. Galaxies of higher star formation rate have higher stellar mass, as reflected in the main--sequence in Figure \ref{Fig:MS}. We also show the feedback-driven turbulence model from \citet{Krumholz2016} for the relation between star formation rate and velocity dispersion, parameterised as SFR$\,\propto$\,$v_{\rm c}^2\sigma^2/Q_{\rm g}$, for different Toomre Q$_{\rm g}$ values, evaluated at the median rotational velocity of the KGES sample, $\langle$\,V$_{\rm 2R_h}$\,$\rangle$\,=\,102\,$\pm$\,8\,km\,s$^{-1}$. The KGES galaxies occupy similar $\sigma_o$\,--\,SFR parameter space as galaxies with Q$_{\rm g}$\,=\,0.25\,--\,3.0.

To quantify the kinematic state of the galaxies in our sample we take the ratio of rotation velocity (V$_{\rm 2R_h}$) to velocity dispersion ($\sigma_0$). Galaxies with dynamics that are dominated by rotation will have V$_{\rm 2R_h}$/$\sigma_0$\,$>$\,1 whilst those with kinematics driven by turbulent pressure-support have V$_{\rm 2.2R_h}$/$\sigma_0$\,$<$\,1. The median ratio of rotation velocity to velocity dispersion in the KGES sample is $\langle$\,V$_{\rm 2R_h}$/$\sigma_0$\,$\rangle$\,=\,1.93\,$\pm$\,0.21 with a 16--84th percentile range of V$_{\rm 2R_h}$/$\sigma_0$\,=\,0.52\,--\,3.89. This is within 1-$\sigma$ of $z$\,$\sim$\,0.9 galaxies in the KROSS survey, which have $\langle$\,V$_{\rm 2R_h}$/$\sigma_0$\,$\rangle$\,=\,2.5\,$\pm$\,1.4 \citep{Harrison2017}, but considerably higher than that \citet{Turner2017} derived for star--forming galaxies at $z$\,$\sim$\,3.5 in the KMOS Deep Survey, with $\langle$\,V$_{\rm 2R_h}$/$\sigma_0$\,$\rangle$\,=\,0.97\,$\pm$\,0.14. This indicates that the kinematics of the galaxies in our sample are, on average, rotation dominated, and representative of the main--sequence population at $z$\,$\sim$\,1.5.

\subsection{Angular Momentum}\label{Sec:AM}

In this section we measure the specific stellar angular momentum (j$_*$) of each galaxy in the KGES sample. We first confirm that the angular momentum of a disc galaxy can be calculated from the integral of the galaxy's one-dimensional rotation and stellar mass profiles as well as from the approximation of asymptotic rotation speed and stellar disc size, as first proposed by \citet{Romanowsky2012} (see also \citealt{Glazebrook2014}). In the following sections, we then explore the correlation of specific stellar angular momentum with stellar mass and analyse the morphological and dynamical properties of the galaxies that scatter about the median j$_*$\,--\,M$_*$ relation.

\subsubsection{Asymptotic and integrated specific stellar angular momentum}

The specific stellar angular momentum is one of most fundamental properties of a galaxy. It combines the rotation velocity profile and the stellar disc size of the galaxy whilst removing the inherent scaling with stellar mass (\citealt{Peebles1969,Fall1980,Fall1983}). 

The specific stellar angular momentum is given by,
\begin{equation}\label{Eqn:AngMom} 
\vec{j_{*}}=\frac{\vec{J_{*}}}{\rm M_{*}}=\frac{\int_{\textbf{r}}(\textbf{r} \times \bar{\textbf{v}}(r) ) \rho_{*}(r) \rm d^3 \textbf{r}}{\int_{\textbf{r}} \rho_{*}(r) \rm d^3 \textbf{r}},
\end{equation}
\noindent where \textbf{r} and $\bar{\textbf{v}}$ are the position and mean-velocity vectors (with respect to the centre of mass of the galaxy) and $\rho(r)$ is the three dimensional density of the stars \citep{Romanowsky2012}. To derive the specific angular momentum from observations, we can use two different approaches which require a number of approximations. We derive the integrated specific stellar angular momentum (j$_*$) of a galaxy by integrating the galaxies rotation velocity and surface brightness profiles. Second, we derive the asymptotic specific stellar angular momentum ($\tilde{\rm j}_{*}$), using the parameterised morphology (e.g. S\'ersic index, stellar continuum size) and asymptotic rotation velocity of the galaxy. In this section we measure both j$_*$ and $\tilde{\rm j}_{*}$ for the galaxies in KGES sample to compare both methods and explore their correlations with galaxy morphology.  In doing so we are assuming that the gas kinematics are good tracers of the stellar angular momentum, which may introduce a small systematic of $\approx$0.1 dex when comparing directly to stellar measurements, based on low--redshift studies (e.g. \citealt{Cortese2014,Cortese2016})

First, we calculate the integrated specific stellar angular momentum (j$_*$) of the KGES galaxies.
If the dynamics of the stars and gas in the galaxies are comprised of only circular orbits, the normal of the specific stellar angular momentum relative to the center of gravity can be written as
\begin{equation}\label{Eqn:AngMom1} 
j_{*}=\left|\frac{\rm \vec{J_*}}{\rm M_{*}}\right|=\frac{\int_{0}^{\infty} \Sigma(r)v(r)r^2 dr}{\int_{0}^{\infty} \Sigma(r) r dr},
\end{equation}
where $\Sigma(r)$ is the azimuthally averaged surface mass density of the stellar component of the galaxy and $v(r)$ is the rotation profile. To evaluate this formula for galaxies in the KGES sample, we use the near-infrared surface brightness profiles I($r$) as a proxy for the surface mass density, under the assumption that mass follows light. As discussed in Section \ref{Sec:Phot} the majority of the galaxies in the sample have $HST$ CANDELS imaging in the near-infrared, that is, rest-frame optical, which traces the old stellar population.

To derive a galaxies surface mass density profile, we calculate the intrinsic surface brightness profile of the galaxy from the $HST$ image and then convolve it with the KMOS PSF. Integrating  this with the rotation velocity profile, measured in Section \ref{Sec:Kin}, we derive a specific stellar angular momentum profile for each galaxy. We then derive an estimate of the total specific angular momentum of each galaxy (j$_{*}$) by extracting the specific stellar angular momentum at 2$\times$ half-stellar mass radius ($\sim$3.4R$_{\rm d}$) from the angular momentum profile. 

The second approach to measuring a galaxy's integrated specific stellar angular momentum (j$_*$) is to derive the galaxy's asymptotic specific stellar angular momentum ($\tilde{\rm j}_{*}$). \citet{Romanowsky2012} showed that the total angular momentum, for galaxies of varying morphological type, can be approximated by a combination of asymptotic rotation speed, stellar disc size and  S\'ersic index,
\begin{equation}\label{Eqn:AngMom2}
\tilde{\rm j}_{*} = k_{n}C_{i}v_{s}R_h,
\end{equation}
\noindent where $v_{\text{s}}$ is the rotation velocity at 2$\times$ the half-light radius (R$_{\text{h}}$), C$_{i}$ is the correction factor  for inclination, assumed to be $\sin^{-1}$($\theta_{inc}$) (see Appendix A of \citealt{Romanowsky2012}) and k$_{n}$ is a numerical coefficient that depends on the S\'ersic index ($n$) of the galaxy and is approximated as:
\begin{equation}\label{Eqn:AngMom3}
k_{n} = 1.15+0.029n+0.062n^2, 
\end{equation}
\noindent This approximation is valid if the surface brightness profile of the galaxy can be well described by a single component  S\'ersic profile parameterised by a half-light radius (R$_{\rm h}$) and  S\'ersic index ($n$). Thus $\Sigma(r) \propto exp(-r/R)$  and assuming the exponential disc is rotating at a constant rotation velocity (v$_s$),
\begin{equation}\label{Eqn:AngMom4}
j_{*}(r) = \left[2+\frac{(r/R)^2}{1+r/R-exp(r/R)}\right]R_h v_s
\end{equation}
\noindent For further details on the potential limitations of this approach we refer the reader to \citet{Glazebrook2014}. 

\begin{figure}
	\centering
	\includegraphics[width=\linewidth]{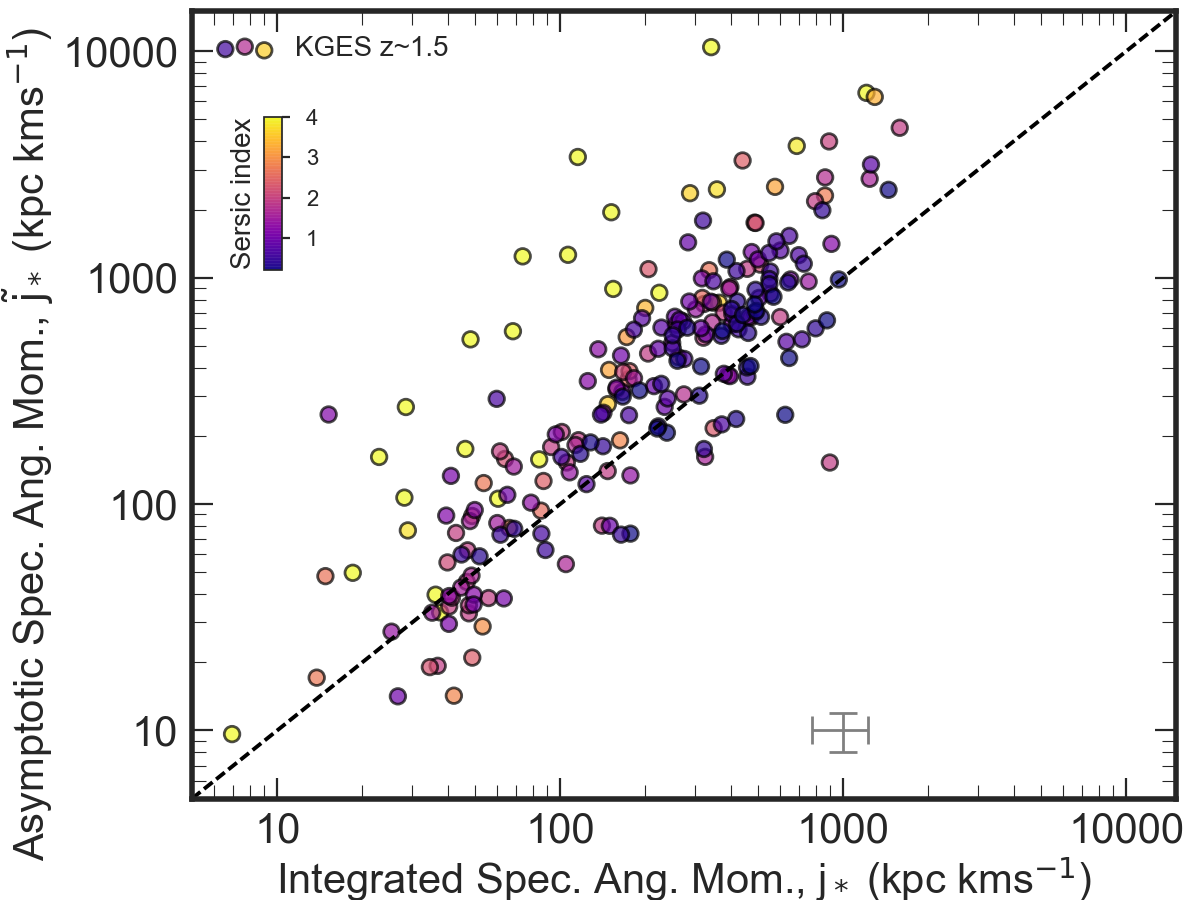}
	\caption{The asymptotic specific stellar angular momentum ($\tilde{\rm j}_{*}$) as a function of the integrated specific angular momentum (j$_{*}$) evaluated at 2$\times$ half-stellar mass radius, for the KGES sample. The black dashed line indicates a one to one relation. The colourbar indicates the S\'ersic index of the galaxy. The scatter below the line is a consequence of deconvolution with a broad--band PSF and convolution with the KMOS PSF. Scatter above the line is driven by galaxies of a higher S\'ersic index in which the integrated specific angular momentum at 2$\times$ half-stellar mass radius is an underestimate of the total angular momentum in the galaxy.}
\label{Fig:jprof_j2rv}
\end{figure}

To compare the two methods, in Figure \ref{Fig:jprof_j2rv} we plot the asymptotic specific stellar angular momentum ($\tilde{\rm j}_{*}$) as a function of the integrated specific angular momentum (j$_{*}$). Galaxies with high S\'ersic index ($n$\,$>$\,2) appear to scatter above the line, with the asymptotic specific angular momentum being over estimated, whilst galaxies with $n$\,$\sim$\,1, scatter about the line.


\begin{figure*}
	\centering
	\includegraphics[width=\linewidth]{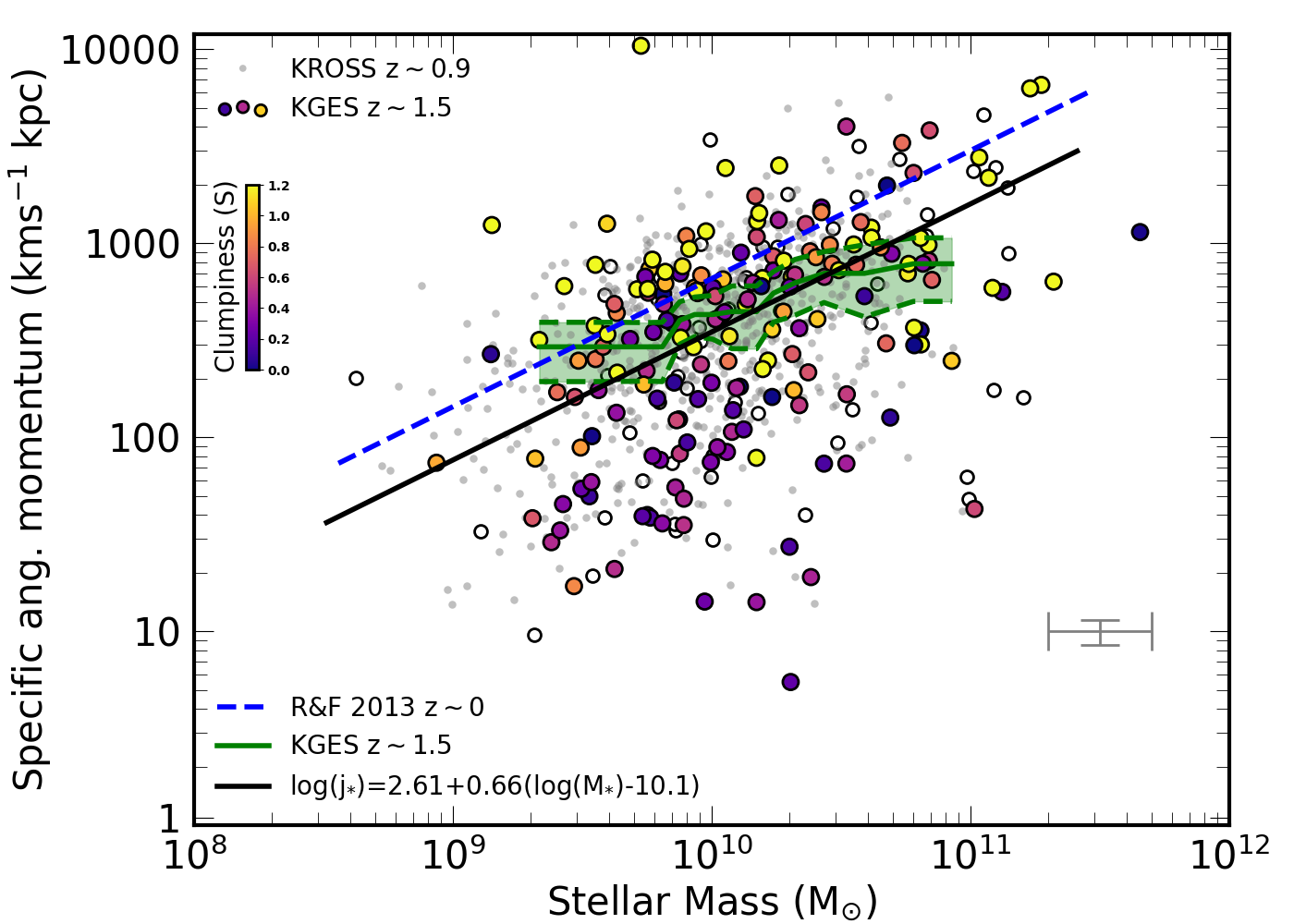}
	\caption{Specific stellar angular momentum as a function of stellar mass. Clumpiness parameter of the KGES  sample shown by the colour map. Lower H$\alpha$ S/N (Quality 3) objects are shown by open circles. KROSS $z$\,$\sim$\,0.9 sample shown as grey points in the background  \citep{Harrison2017}.
A parametric fit to the disc component of  z\,$\sim$\,0 galaxies as derived by \citet{RF2013}  is shown by the blue line.
The green shaded region and dashed lines indicate the median trend of the KGES galaxies and their 1$\sigma$ scatter. The black line is a fit to the KGES data of the form $\log_{10}(j_*)$\,=\,$\alpha$\,+$\,\beta\,(\log_{10}(\rm M_*/M_{\odot})-10.10)$, with the slope fixed to $\beta$\,=\,0.66 and a derived intercept of $\alpha$\,=\,2.61.
	The KGES sample occupy a similar region of parameter space to KROSS but offset to lower angular momentum for given stellar mass than \citet{RF2013} z\,$\sim$\,0 pure disc sample. The galaxies show a trend of increasing specific angular momentum with stellar mass whilst having a broad range of specifc stellar angular momentum at fixed stellar mass that correlates with the Clumpiness of the galaxy.}
\label{Fig:JM_Clump}
\end{figure*}

To understand the source of the scatter within this plane we measure both the asymptotic and integrated specific angular momentum for 1000 mock galaxies with $\log$(M$_{*}$[M$_{\odot}$])\,=\,9\,--\,10.5, S\'ersic index $n$\,=\,0.5\,--\,8 and half stellar mass radii in the range R$_{\rm h}$\,=\,0\farcs{1}\,--\,2\farcs{0}. A tight correlation between $\tilde{\rm j}_{*}$ and j$_{*}$ is identified for galaxies with $n$\,=\,0.5\,--\,2 of all stellar masses and continuum sizes, with $\langle$\,$\tilde{\rm j}_{*}$/j$_{*}$\,$\rangle$\,=\,0.88\,$\pm$\,0.03, when the PSF of both the mock broad--band and integral field data is $\approx$0 arcseconds.  
The integrated specific stellar angular momentum (j$_{*}$) overestimates the angular momentum of galaxies, when a non-zero PSF is used. The inner regions of the angular momentum profile of the galaxy are not resolved in the convolution process, especially when the PSF is comparable to the galaxies' stellar continuum size.


For mock galaxies with S\'ersic index $n$\,=\,2\,--\,8,  $\langle$\,$\tilde{\rm j}_{*}$/j$_{*}$\,$\rangle$\,=\,2.88\,$\pm$\,0.94 with the integrated specific stellar angular momentum being underestimated in galaxies of a higher S\'ersic index. \citet{Romanowsky2012} comment that the reliability of $\tilde{\rm j}_{*}$ $\approx$ j$_{*}$ depends systematically on the density profile, where for galaxies with $n$\,=\,2, 4, and 6, $\tilde{\rm j}_{*}$\,=\,j$_{*}$  at R\,$\sim$\,2R$_{\rm h}$, 4.5R$_{\rm h}$, and 10R$_{\rm h}$, highlighting that the extended envelopes of higher S\'ersic index galaxies contribute more to j$_{*}$.

For the remainder of the analysis on the KGES sample we therefore adopt $\tilde{\rm j}_{*}$ (Equation \ref{Eqn:AngMom2}) as the estimate of the total specific stellar angular momentum in the galaxies which is expected to recover the total angular momentum of a galaxy to within four per cent  \citep{Romanowsky2012}.

\subsection{Summary of Morphological and Dynamical Properties}

We detected H$\alpha$ and [N{\sc{ii}}] emission in 243 of our targets (84 per cent of the sample) and showed that they are representative of `main--sequence' star--forming galaxies at $z$\,$\sim$\,1.5 (Section \ref{Sec:MS}). We parameterised their rest-frame optical morphology of this sample of spatially resolved galaxies, both parametrically, identifying on average their stellar light distributions follow an exponential disc with a median S\'ersic index of $\langle$\,$n$\,$\rangle$\,=\,1.37\,$\pm$\,0.12 (Section \ref{Sec:Galfit}), and non--parametrically, showing that the galaxies in the KGES sample have symmetrical and clumpy morphologies (Section \ref{Sec:CAS}).

Exploiting the KMOS observations, we showed the kinematics of the KGES galaxies align with that of star--forming discs with well defined ordered rotation (Figure \ref{Fig:kin_pg0}) with a median rotational velocity of $\langle$\,V$_{\rm 2R_h}$\,$\rangle$\,=\,102\,$\pm$\,8\,km\,s$^{-1}$. A full catalogue of all observable properties measured from the KGES galaxies will be published in Tiley et al. (in prep.). 
In the following sections we use these observed properties of the KGES galaxies to analyse more derived quantities, (e.g. specific angular momentum) and explore the connection between a galaxy's gas dynamics and rest--frame optical morphology.

\section{Discussion}\label{Sec:Diss}

\subsection{The Specific Angular Momentum of gas discs at $z$\,$\sim$\,1.5}


The correlation between specific stellar angular momentum and stellar mass is well established at $z$\,$\sim$\,0 \citep[e.g.][]{Fall1980,Posti2018} with higher stellar mass galaxies having higher specific angular momentum according to a scaling j$_*$\,$\propto$\,M$_*^{2/3}$ \citep[e.g.][]{Fall1983,Mo1998}. \citet{Romanowsky2012} updated the work by \citealt{Fall1983} with new observations of galaxies spanning a range of  morphologies, confirming that for a fixed stellar mass, galaxy discs have a factor 5-6$\times$ more angular momentum than spheroidal galaxies.

In Figure \ref{Fig:JM_Clump} we plot the specific stellar angular momentum of the KGES sample as a function of their stellar mass. The median specific stellar angular momentum in the sample is $\langle$\,j$_{*}$\,$\rangle$\,=\,391\,$\pm$\,53 km\,s$^{-1}$\,kpc with a 16-84th percentile range of  j$_*$\,=\,74\,--\,1085 km\,s$^{-1}$\,kpc. To place the KGES sample in context with the j$_*$\,--\,M$_*$ plane, we compare the specific stellar angular momentum to other surveys of star--forming galaxies across a range of redshift. We include the \citet{RF2013} pure disc sample of star--forming $z$\,$\sim$\,0 galaxies as well the KROSS (\citealt{Harrison2017}) $z$\,$\sim$\,0.9 sample. On average, for a given stellar mass, KGES galaxies occupy a similar  region of parameter space to the KROSS sample whilst being offset to lower specific stellar angular momentum than the \citet{RF2013} $z$\,$\sim$\,0 sample. 
It should be noted that other studies have also suggested minimal evolution in the zero-point offset in the j$_*$\,--\,M$_*$ from $z$\,$\sim$\,1 to $z$\,$\sim$\,0 \citep[e.g][]{Marasco2019}.

To quantify the specific stellar angular momentum and stellar mass plane in the KGES sample,
we fit a relation of the form $\log_{10}(j_*$)\,=\,$\alpha$\,+\,$\beta$\,(log$_{10}$(M$_*$/M$_{\odot}$)\,--\,10.10). At low redshift the relationship between galaxy and halo angular momentum is approximated by j$_*$/j$_{\rm halo}$\,$\propto$\,$\rm \left(M_*/M_{halo}\right)^{2/3}$ \citep[e.g][]{Romanowsky2012,Obreschkow2015,Fall2018,Sweet2019,Posti2019}. A power law index of $\beta$\,=\,0.66
at high--redshift implies that dark matter haloes in a $\Lambda$CDM Universe are scale free. However, the stellar mass fraction (M$_*$/M$_{\rm halo}$) varies strongly with halo mass, \citep[e.g.][]{Behroozi2019, Sharma2019} and therefore it is not clear that the exponent should also hold for stars. To test whether this scaling holds in high--redshift galaxies, we fit the $j_*$\,--\,M$_*$ plane 
using a  chi-squared minimisation to find the best fit parameters of the linear model.
For the KGES  galaxies, with an unconstrained fit, we derive a slope of $\beta$\,=\,0.53\,$\pm$\,0.10  with a normalisation of $\alpha$\,=\,2.63\,$\pm$\,0.04 

The slope of the $j_*$\,--\,M$_*$ plane is consistent within 1.3-$\sigma$ of that derived from the assumption j$_*$/j$_{\rm halo}$\,$\propto$\,$\rm \left(M_*/M_{halo}\right)^{2/3}$. 
Given this similarity for the following analysis we make the assumption and fix $\beta$\,=\,0.66 (i.e assuming j$_*$/j$_{\rm halo}$\,$\propto$\,$\rm \left(M_*/M_{halo}\right)^{2/3}$), which allows comparison to lower redshift surveys \cite[e.g][]{RF2013}.
We re-fit the $j_*$\,--\,M$_*$ plane, constraining the slope to be $\beta$\,=\,0.66
and derive a normalisation $\alpha$\,=\,2.60\,$\pm$\,0.03 for all 235 spatially resolved KGES galaxies.
We note that the parameterisation of the $j_*$\,--\,M$_*$ plane is dependent on the uncertainties on the stellar mass which can be significant \citep[e.g.][]{Pforr2012}. We have adopted a conservative $\pm$\,0.2\,dex uncertainty as demonstrated by \citet{Mobasher2015} to account for systematic effects.

Across the whole sample of targeted 288 KGES galaxies, there is a range of H$\alpha$ signal to noise, with some galaxies having very low signal to noise kinematics and rotation curves. Subsequently, dynamical measurements of these galaxies are more uncertain. To understand the effect these lower quality targets have on our analysis, we define four quality flags.with the following kinematic criteria that is based on the signal to noise of the galaxy integrated H$\alpha$ emission and the extrapolation of the observed rotation curve.:
\begin{itemize}
\item Quality 1: H$\alpha$\,$>$\,50 S/N and R$_{\rm last}$/2R$_{\rm h}$\,$>$\,1
 \item Quality 2: 20$<$\,H$\alpha$ S/N\,$<$\,50 and R$_{\rm last}$/2R$_{\rm h}$\,$>$\,1
 \item Quality 3: H$\alpha$ S/N $<$\,20 or 0.3$<$\,R$_{\rm last}$/2R$_{\rm h}$\,$<$\,1.5
    \item Quality 4: H$\alpha$ S/N $<$\,1 or R$_{\rm last}$/2R$_{\rm h}$\,$<$\,0.1
\end{itemize}


Of the 288 galaxies, 201 are classified as either quality 1 (107 galaxies) or quality 2 (94 galaxies). 42 galaxies are labelled as quality 3 whilst 45 galaxies have the lowest quality kinematic and broadband data and are labelled quality 4. If we fit $\log_{10}(j_*$)\,=\,$\alpha$\,+\,$\beta$\,(log$_{10}$(M$_*$/M$_{\odot}$)\,--\,10.10) to just quality 1 $\&$ 2 galaxies we establish a normalisation of $\alpha$\,=\,2.61, indicating that including only high quality targets gives the same normalisation as the full sample.


\begin{figure*}
 	\includegraphics[width=1\linewidth,trim={2cm 0cm 2cm 0}]{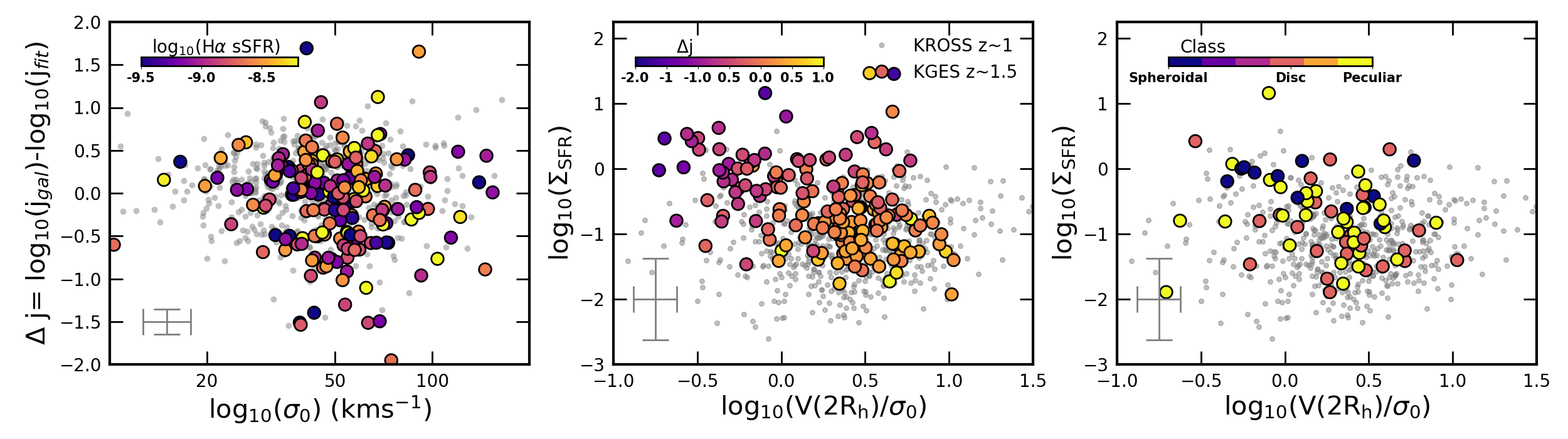}
	\caption{Left panel: The angular momentum offset from the parametric fit $\log_{10}(j_*)$\,=\,2.61\,+$\,0.66\,(\log_{10}(\rm M_*/M_{\odot})-10.10)$ ($\Delta$j) as function of velocity dispersion ($\sigma_0$) coloured by the H$\alpha$ specific star formation rate. We identify no correlation between a galaxies position in the $j_*$\,--\,M$_*$ plane and the velocity dispersion or H$\alpha$ specific star formation rate (e.g. turbulence of the interstellar medium) of the galaxy. Middle and Right panel: The H$\alpha$ star formation rate surface density ($\Sigma_{\rm SFR}$) as a function of the ratio of rotation velocity to velocity dispersion (V(2R$_{\rm h}$)/$\sigma_0$). The middle panel is coloured by $\Delta$j, whilst the right panel is coloured by visual morphological class, as defined in Section \ref{Morph_j}. In all three panels the KROSS $z$\,$\sim$\,0.9 sample is shown by the grey points. The median uncertanity is shown in the lower left corner of each panel. Galaxies of higher $\Sigma_{\rm SFR}$, are more dispersion dominated, with lower specific stellar angular momentum, resembling more spheroidal morphologies. Disc galaxies have lower $\Sigma_{\rm SFR}$, are more rotation dominated, and have higher specific stellar angular momentum whilst peculiar galaxies tend to have high 
	$\Sigma_{\rm SFR}$ 	whilst being rotation dominated, with high specific stellar angular momentum.}	
	
\label{Fig:deltaj_dyn}
\end{figure*}

\begin{figure*}
	\centering
	\includegraphics[width=\linewidth]{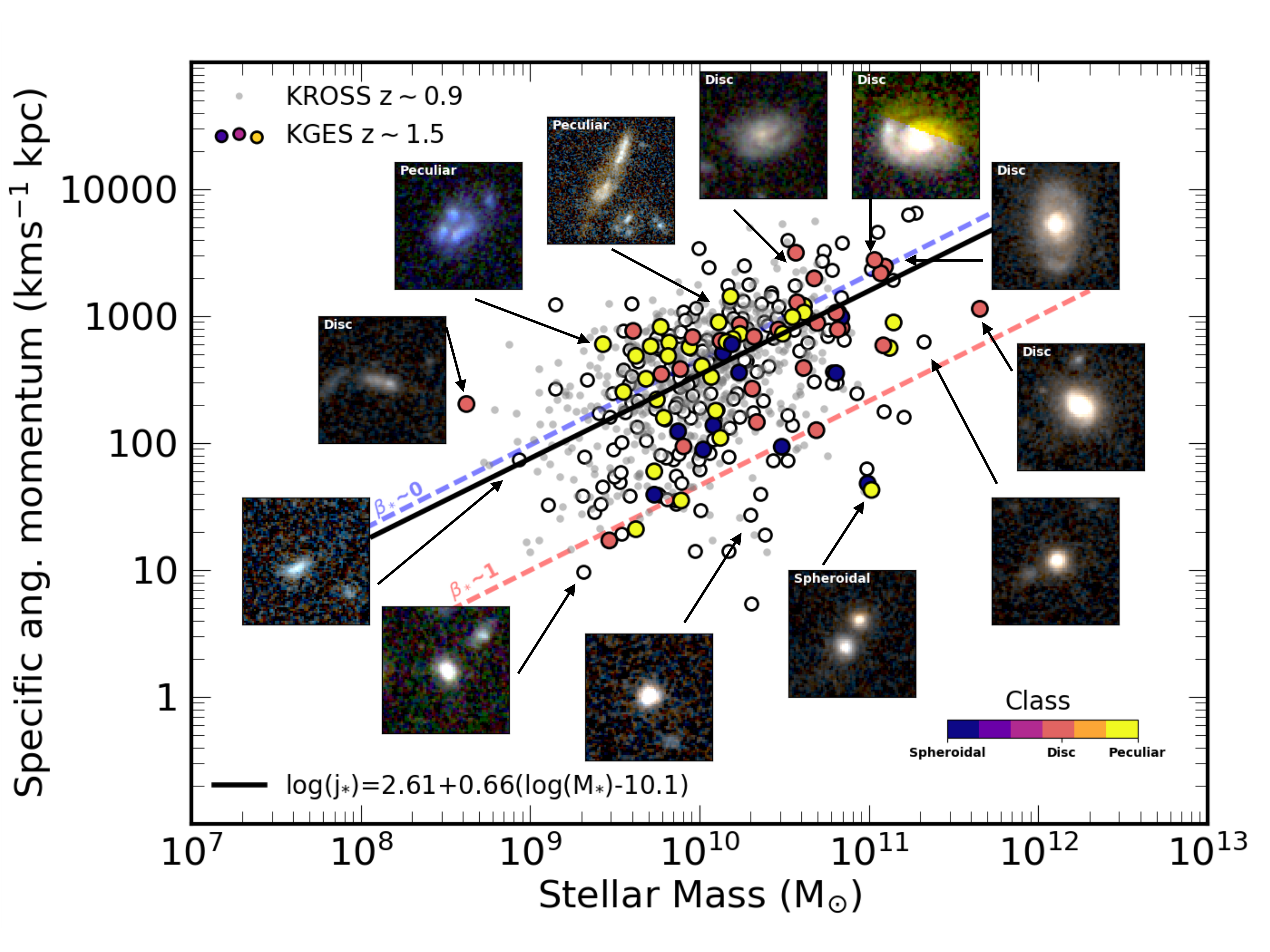}
	\caption{Specific stellar angular momentum as a function of stellar mass. Visual morphology of the KGES  sample shown by the colour map. Quality 3 and 4 objects shown by open circles. KROSS $z$\,$\sim$\,0.9 sample shown as grey points in the background \citet{Harrison2017}. The black line is a fit to the KGES data of the form $\log_{10}(j_*)$\,=\,$\alpha$\,+$\,\beta\,(\log_{10}(\rm M_*/M_{\odot})-10.10)$, with the slope fixed to $\beta$\,=\,0.66 and a derived intercept of $\alpha$\,=\,2.61. Fixed stellar bulge to total ratio ($\beta_*$) lines from \citet{Romanowsky2012} are shown by the blue and red lines. $HST$ wide field camera  colour images of some of the galaxies are shown around the edge of the figure with the visual class of the galaxy indicated. There is a clear correlation between the position of the galaxy in the specific stellar angular momentum stellar mass plane and the galaxies visual morphology.}
\label{Fig:JM_Class}
\end{figure*}

\subsection{Dynamics and Angular Momentum}\label{Sec:AM_corr}

With a  sample of 235 galaxies with spatially resolved gas kinematics we can investigate the scatter about the median j$_*$\,--\,M$_*$ trend that is driven by physical processes in a galaxy's evolution. In this section we explore how the scatter correlates with the galaxy's dynamical properties (e.g. rotation velocity, turbulence, star formation rate surface density). 

To quantify the position of a galaxy in the j$_*$--M$_*$ plane we define the parameter, $\Delta$j as $\Delta$j\,=\,$\rm log_{10}(j_{gal})$\,--\,$\rm log_{10}(j_{fit})$. Where j$\rm_{gal}$ is the specific stellar angular momentum of the galaxy and j$\rm_{fit}$ is the specific stellar angular momentum of the parametric fit to the survey at the same stellar mass (see \citet{Romanowsky2012} Equation 12). Galaxies that lie above the parametric fit of the form $\log_{10}(j_*)$\,=\,2.61\,+$\,0.66\,(\log_{10}(\rm M_*/M_{\odot})-10.10)$ will have positive $\Delta$j whilst those galaxies that lie below the line will have negative $\Delta$j values.


In Figure \ref{Fig:deltaj_dyn} we show the correlation between velocity dispersion ($\sigma_0$) and $\Delta$j, with the galaxies coloured by their H$\alpha$ specific star formation rate. The KROSS $z$\,$\sim$\,0.9 sample is shown for comparison. We identify a no correlation between velocity dispersion and $\Delta$j, with a spearman rank coefficient of $r$\,=\,$-$0.09. This indicates that galaxies of higher angular momentum do not necessarily have less turbulence and thinner discs. This appears to be the case at both $z$\,$\sim$\,0.9 and $z$\,$\sim$\,1.5. We have also identified no significant correlation between the H$\alpha$ specific star formation rate and $\Delta$j of KGES galaxies
indicating that more turbulent galaxies with higher specific star formation rates do not necessarily have lower specific angular momentum.  


In Figure \ref{Fig:deltaj_dyn} we also show the star formation rate surface density ($\Sigma_{\rm SFR}$) as a function of the ratio of rotation velocity to velocity dispersion (V(2R$_{\rm h}$)/$\sigma_0$) for both KGES and KROSS samples, identifying a spearman rank coefficient of $r$\,=\,$-$0.42.
Galaxies that are dispersion dominated (low V(2R$_{\rm h}$)/$\sigma_0$), tend to have higher $\Sigma_{\rm SFR}$, and low specific angular momentum (negative $\Delta$j).

\begin{figure*}
	\includegraphics[width=1\linewidth,trim={5cm 2cm 5cm 1cm}]{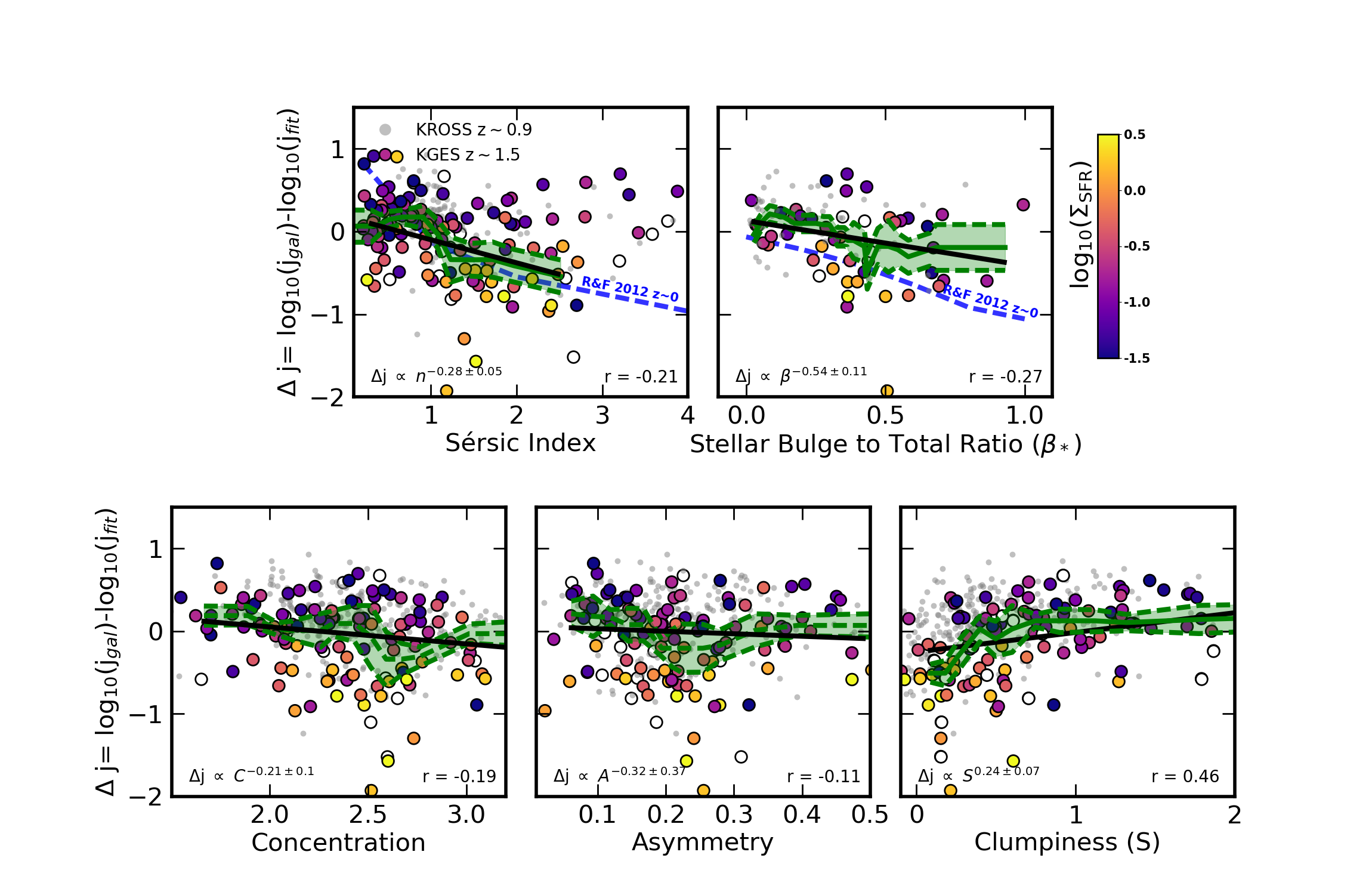}
	\caption{The angular momentum offset from the parametric fit $\log_{10}(j_*)$\,=\,2.61\,+$\,0.66\,(\log_{10}(\rm M_*/M_{\odot})-10.10)$ ($\Delta$j) as function of  S\'ersic index, stellar bulge to total ratio ($\beta_{*}$), Clumpiness, Asymmetry, and Concentration for the KGES galaxies measured in the CANDELS F814W $HST$ band. Open circles show  quality 3 $\&$ 4 galaxies, whilst quality 1 $\&$ 2 galaxies are coloured by their H$\alpha$ star formation rate surface density ($\Sigma_{\rm SFR}$). In the top two panels we show a $z$\,$\sim$\,0 comparison sample from \citet{Romanowsky2012}. The KROSS survey is shown by the grey points in the background, with $\Delta$j measured relative to the parametric fit to the KROSS galaxies. The green line and shaded region indicates a running median and 1\,$\sigma$ error to the KGES quality 1 $\&$ 2 galaxies, and the black line is a parametric fit. 
	Galaxies in the KGES sample with high specific angular momentum for a given stellar mass, on average have lower S\'ersic index and stellar bulge to total ratio whilst being more clumpy and asymmetrical.}
\label{Fig:deltaj_morph}
\end{figure*}

\subsection{Morphology and Angular Momentum}\label{Morph_j}

Now that we have explored the connection between a galaxy's dynamics and its specific angular momentum, identifying galaxies that are more rotation dominated generally have higher angular momentum and lower star-formation rate surface densities,
we now explore the connection to the galaxy's parameterised rest--frame optical morphology.

In the local Universe strong correlations have been identified at fixed stellar mass between a galaxy's S\'ersic index, stellar bulge to total ratio and specific angular momentum.
Both \citet{Romanowsky2012} and \citet{Cortese2016} identified that the more bulge dominated, spheroidal, a system is, the lower its specific angular momentum for a given stellar mass will be. The scatter about the $j_*$\,--\,M$_*$ plane at low redshift is driven by the variation in
S\'ersic index and stellar bulge to total ratio of the galaxies \citep[e.g][]{Glazebrook2014,Fall2018,Sweet2018}.

As as first approach, we adopt the visual classifications of galaxy morphology from \citet{Huertas-Company2015}, who use convolutional neural networks to categorize the $HST$ F160W morphology of 50,000 galaxies in the CANDELS survey. By training the algorithm on the GOOD-S CANDELS field, which has been previously visually classified by \citet{Kartaltepe2016}, \citet{Huertas-Company2015} were able to accurately classify a galaxies morphology with a 1 per cent mis-classification. We cross match the KGES survey in the overlapping region with galaxies in the \citet{Huertas-Company2015} sample, identifying 122 galaxies. Of which, 84 galaxies have a visual classification as either spheroidal, disc or peculiar morphology. The remaining 34 galaxies were not definitively classified by the neural network.


In Figure \ref{Fig:deltaj_dyn} we show the relation between star formation rate surface density ($\Sigma_{\rm SFR}$) and the ratio of rotation velocity to velocity dispersion (V(2R$_{\rm h}$)/$\sigma_0$), with KGES galaxies coloured by their visual morphologies. More dispersion dominated galaxies with higher $\Sigma_{\rm SFR}$ tend to be the more spheroidal with $\langle$\,V(2R$_{\rm h}$)/$\sigma_0$\,$_{\rm spheroidal}$\,$\rangle$\,=\,1.19\,$\pm$\,0.68. Rotation dominated KGES galaxies (high V(2R$_{\rm h}$)/$\sigma_0$), tend to have lower $\Sigma_{\rm SFR}$ with high specific angular momentum, and have visual morphologies that appear as either discs or peculiar systems with $\langle$\,V(2R$_{\rm h}$)/$\sigma_0$\,$_{\rm disc}$\,$\rangle$\,=\,2.33\,$\pm$\,0.40. whilst $\langle$\,V(2R$_{\rm h}$)/$\sigma_0$\,$_{\rm peculiar}$\,$\rangle$\,=\,2.22\,$\pm$\,0.37.

To understand this link between morphology and angular momentum further, we show the specific stellar angular momentum stellar mass plane for the KGES survey, in Figure \ref{Fig:JM_Class}, with galaxies coloured by their `visual morphology'. 
Galaxies classified as spheroidal  appear to lie clearly below the fit, as expected due to their smaller stellar continuum sizes, whilst galaxies labelled as discs appear to lie above the fit. Galaxies labelled as peculiar appear to be scattered about the best fit line highlighting the diversity of the peculiar galaxies morphology and kinematic state.

For galaxies scattered about the median trend, in the specific stellar angular momentum stellar mass plane, in Figure \ref{Fig:JM_Class}, we show the $HST$ wide field camera colour images. For a given stellar mass, those galaxies that have the highest angular momentum have more prominent discs with the presence of spiral arms. Whilst galaxies with the lowest angular momentum are much more spheroidal and spheroidal, as expected. We note however, that the spheroidal galaxies may appear to have low angular momentum because their rotation is unresolved in the KMOS  observations. The higher stellar mass, high angular momentum KGES galaxies show strong signs of significant bulge components in their colour images. This is in agreement with the evolution of stellar mass and stellar bulge-to-total ratio identified in both simulations (e.g. \citealt{Trayford2018}) and observations (e.g. \citealt{Gillman2019}).

\subsubsection{Quantised Morphology and Dynamics}

To interpret this connection between morphology and angular momentum further, we explore the correlation between a galaxy's position in the j$_*$\,--\,M$_*$ plane and its quantised (both parametric and non-parametric) morphology as derived in Section \ref{Sec:Galfit}.
In Figure \ref{Fig:deltaj_morph} we plot $\Delta$j as function of S\'ersic index, stellar bulge to total ratio ($\beta_{*}$), Clumpiness, Asymmetry, and Concentration for KGES galaxies with CANDELS F814W $HST$ imaging. We select this subsample of KGES galaxies with the highest quality data, to allow accurate comparison between the integrated parametric and non-parametric measures of morphology. 

The S\'ersic index of KGES galaxies has a weak negative correlation with a galaxy's position in the j$_*$\,--\,M$_*$ plane, of the form $\Delta$j\,$\propto$\,$n^{-0.27\,\pm\,0.05}$  with a spearman rank coefficent of $r$\,=\,-0.20,
and this weakens slightly with the inclusion of galaxies from KROSS. 
Galaxies of higher S\'ersic index at $z$\,$\sim$\,1.5  have lower $\Delta$j and this appears to be less common at $z$\,$\sim$\,0.9. We show the relation between $\Delta$j and S\'ersic index for $z$\,$\sim$\,0 galaxies from \citet{Romanowsky2012}. The parameterisation of the relation is taken from \citet{Cortese2016} who established the j$_*$\,--\,M$_*$\,--\,$n$ relation for the SAMI survey. We note the parameterisation derived in \citet{Cortese2016} is for a morphologically diverse population of both quiescent and star--forming low redshift galaxies, and therefore should not be compared directly to our sample of star--forming selected high--redshift galaxies.
The relation between stellar mass, S\'ersic index and specific angular momentum  can be parameterised as,
\begin{equation}
\rm log(j/kpc\,km\,s^{-1})\,=\,a\,\times\,log(M_{*}/M_{\odot})\,+\,b\,\times\,log(n)\,+\,c
\end{equation}
\noindent where $a$\,=\,1.05, $b$\,=\,$-$1.38 and c\,$\,=\,$\,$-$8.18. Using the sample of $z$\,$\sim$\,0 galaxies presented in \citet{Romanowsky2012},
we establish the relation between $\Delta$j and S\'ersic index for $z$\,$\sim$\,0 galaxies indicated by the dashed line in Figure \ref{Fig:deltaj_morph}. The relation is very similar to that identified in the KGES sample at $z$\,$\sim$\,1.5, with higher S\'ersic index galaxies having lower specific angular momentum.

The stellar bulge to total ratios ($\beta_{*}$) for both KROSS and KGES galaxies are taken from \citet{Dimauro2018} who derive $\beta_{*}$ using a multi-wavelength machine learning algorithm for $\sim$\,18,000  galaxies in the $HST$ CANDELS field selected to have an F160W magnitude of $<$23 in the redshift range $z$\,=\,0\,--\,2. In Figure \ref{Fig:deltaj_morph} we plot $\Delta$j as a function of $\beta_{*}$, derived from only F160W $HST$ imaging, and identify a moderate negative correlation of $\Delta$j\,$\propto$\,$\beta_*^{-0.27\,\pm\,0.36}$ and a spearman rank coefficent of $r$\,=\,-0.27,
with lower angular momentum galaxies having higher bulge to total ratios. A similar correlation is present in KROSS at $z$\,$\sim$\,0.9, and when the two surveys are combined we derive $\Delta$j\,$\propto$\,$\beta_*^{-0.51\,\pm\,0.18}$.
This is in agreement with the correlation between $\Delta$j and $n$, with higher S\'ersic index stellar light distributions corresponding to more bulge dominated systems.

\begin{figure*}
	\includegraphics[width=1\linewidth,trim={5cm 2cm 3cm 1cm}]{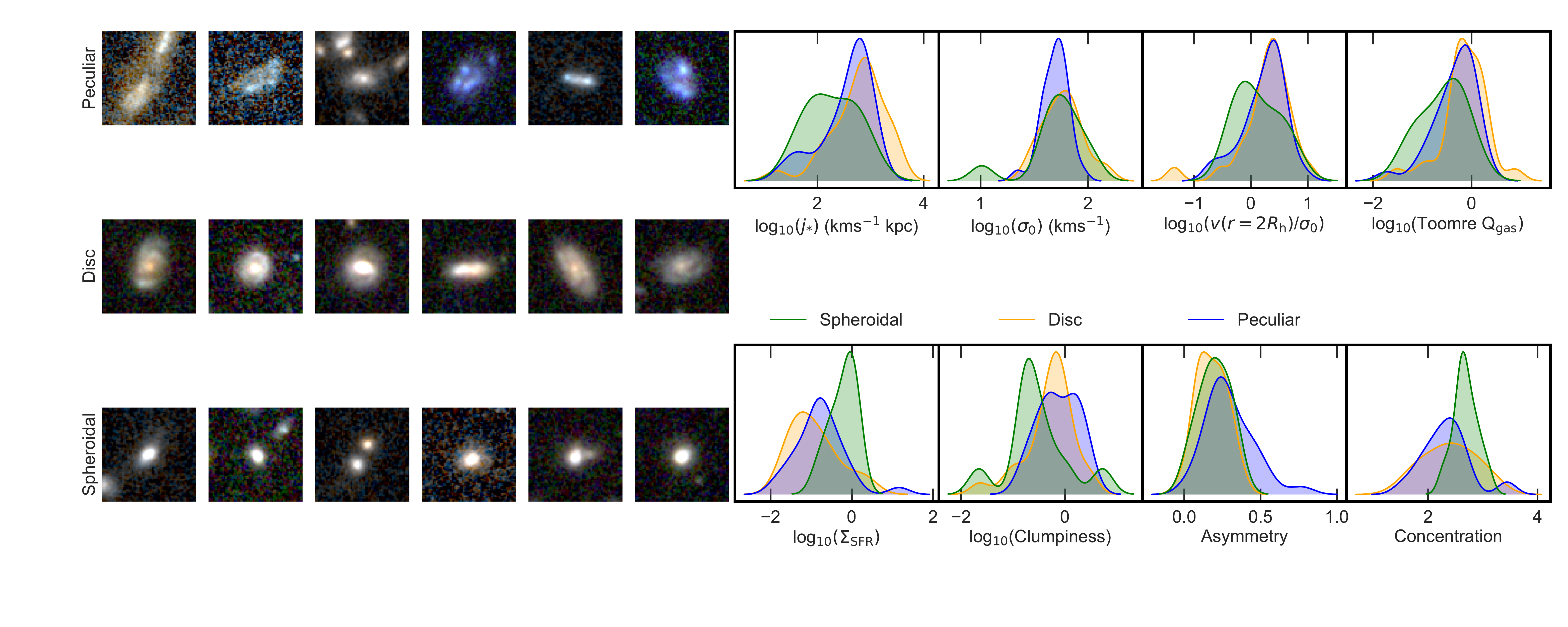}
	\caption{HST colour images of KGES galaxies Spheroidal, Disc and Peculiar morphological classes (left) with the kernel density distribution of specific angular momentum (j$_*$), velocity dispersion ($\sigma_0$), ratio of rotation velocity to velocity dispersion (V(2R$_{\rm h}$)/$\sigma_0$), disc stability (Q$_{\rm gas}$), H$\alpha$ star formation rate surface density ($\Sigma_{\rm SFR}$), Clumpiness, Asymmetry and Concentration (right). The velocity dispersion and concentration of Spheroidal, Disc and Peculiar galaxies are very similar. Spheroidal galaxies have lower specific stellar angular momentum, are more dispersion dominated, have lower Toomre $Q_{\rm gas}$, are less clumpy, more asymmetrical but have higher $\Sigma_{\rm SFR}$ than Disc--like galaxies. Peculiar galaxies on average have the same specific stellar angular momentum, are similarly rotation dominated, but have lower Toomre Q$_{\rm gas}$ and are more  clumpy, more asymmetrical but with higher $\Sigma_{\rm SFR}$ than Disc--like galaxies.}
\label{Fig:morph_hist}
\end{figure*}


\citet{Fall2018} identify a strong correlation between a galaxy's position in the specific stellar angular momentum stellar mass plane and stellar bulge to total ratio in a sample of local galaxies. 
Galaxies with fixed bulge to total ratio follow parallel tracks in the j$_*$\,--\,M$_*$ plane, with $\beta_{*}$\,$\sim$0 (Sc, Sb) galaxies having the highest normalisation and $\beta_{*}$\,$\sim$1 (E) galaxies having the lowest (Figure \ref{Fig:JM_Class}). They conclude that the j$_*$\,--\,M$_*$\,--\,$\beta_{*}$ scaling provides an alternative to the Hubble classification of galaxy morphology. In Figure \ref{Fig:deltaj_morph}, we plot the correlation between $\Delta$j and bulge to total ratio derived from the relations and galaxies presented in \citet{Romanowsky2012}. The $z$\,$\sim$\,0 relation is offset to lower angular momentum than our $z$\,$\sim$\,1.5 sample, with more bulge dominated galaxies having lower angular momentum, than a galaxy with the same $\beta_{*}$ at $z$\,$\sim$\,1.5. We note the scatter in the $\Delta$j\,--\,$\beta_{*}$ and $\Delta$j\,--\,$n$ plane maybe driven by a combination of resolution effects, whereby we do not resolve the rotation in spheroidal objects, nor do we resolve the kinematics on sub--kpc scales revealing potential merging kinematic components. Equally the galaxy population may contain a number of massive early-type galaxies with evolved bulges that have high S\'ersic index and bulge to total ratios as well as the dominant population of spheroidal star-forming galaxies that have a high central star--formation rates.

The position of a galaxy in the j$_*$\,--\,M$_*$ shows a weak negative correlation with the Concentration of the galaxy's stellar light with $\Delta$j\,$\propto$\,$C^{\,-0.2\,\pm\,0.1}$ ($r$\,=\,-0.18). This is as expected as more concentrated galaxies have higher S\'ersic indicies and higher bulge to total ratios. The asymmetry of the galaxy however shows no significant correlation, with $\Delta$j\,$\propto$\,$A^{\,-0.32\,\pm\,0.37}$ and a spearman rank coefficent of $r$\,=\,-0.11. The Clumpiness of the light distribution however indicates a moderate positive trend ($\Delta$j\,$\propto$\,$S^{\,0.24\,\pm\,0.07}$) with $\Delta$j with a spearman rank coefficent of $r$\,=\,-0.46.
This indicates galaxies that are more clumpy  and less concentrated have higher angular momentum than the average galaxy in the survey for a given stellar mass, regardless of the asymmetry of the light profile. The correlation with the symmetry of the galaxy is less well constrained due to the large uncertanity on the exponent. As shown in Figure \ref{Fig:deltaj_morph}, galaxies with higher star formation rate surface density have lower specific angular momentum at fixed stellar mass.


We infer that the correlations in  Figures \ref{Fig:deltaj_dyn} $\&$ \ref{Fig:deltaj_morph} could be driven by spheroidal objects with low angular momentum being very concentrated and smooth, whilst high angular momentum disc galaxies with spiral arms and significant bulge components are more clumpy and but have similar levels of asymmetry. Peculiar galaxies in the KGES sample also are very clumpy and asymmetrical but still maintain high specific angular angular momentum. 

\subsubsection{Qualitative Morphology and Dynamics}

As shown in Figure \ref{Fig:deltaj_morph}, high specific angular momentum galaxies tend to have higher clumpiness and are less bulge dominated with lower S\'ersic indices. Figure \ref{Fig:JM_Class} shows that high angular momentum galaxies generally have disc dominated or peculiar morphologies. Using the visual classifications established from \citet{Huertas-Company2015}, the medium clumpiness of peculiar galaxies in the KGES sample is $\langle$\,S$_{\rm peculiar}$\,$\rangle$\,=\,0.70\,$\pm$\,0.27 whilst for disc galaxies $\langle$\,S$_{\rm disc}$\,$\rangle$\,=\,0.58\,$\pm$\,0.10. The S\'ersic index of peculiar systems is $\langle$\,$n_{\rm peculiar}$\,$\rangle$\,=\,0.88\,$\pm$\,0.14 whilst disc galaxies have a medium value of $\langle$\,$n_{\rm disc}$\,$\rangle$\,=\,1.19\,$\pm$\,0.28. The quantitative, parametric and non-parametric, measures of a galaxies morphology are successful in isolating spheroidal systems however they are less reliable in distinguishing peculiar galaxies from disc--dominated ones. Consequently, we next focus on the dynamical differences between the visual morphological classes in the KGES survey.


Before we compare the kinematic properties of galaxies with different morphologies, we first infer an approximation for the stability of the gas disc in each galaxy. To analyse the interplay between the rotational velocity, velocity dispersion and star formation rate surface density, we quantify the average stability of the galactic disc in each galaxy against local gravitational collapse, as parameterised by the Toomre stability parameter.

From the Jeans criterion, a uniform density gas cloud will collapse if its self-gravity can overcome the internal gas pressure \citep{Jeans1902}. However in  a galactic disc the differential rotation of the galaxy provides additional support to the internal gas pressure of the gas cloud. If the gas cloud becomes too large it will be torn apart by shear, faster than the gravitational free fall time \citep{Toomre1964}. For a thin gas disc, this stability criterion of the balance between shear, pressure support and self-gravity can be quantified by the Toomre Q$_{\rm gas}$ parameter which is defined as,
\begin{equation}
\rm Q_{gas}\,=\,\frac{\sigma_{gas}\kappa}{\pi G \Sigma_{gas}},
\end{equation}
\noindent where $\sigma_{\rm gas}$ is the line-of-sight velocity dispersion, $\Sigma_{\rm gas}$ is the gas surface density of the disc and $\kappa$ is the epicyclic frequency of the galaxy and is approximated as $\rm \kappa=aV/R$. Within which V is the rotational velocity of the disc at radius R and a\,=\,$\sqrt{2}$ for a flat rotation curve. The rotational velocity and velocity dispersion are measured at 2R$_{\rm h}$ from the kinematic profiles of each galaxy (Secion \ref{Sec:Kin}).

We use the Kennicutt-Schmidt (KS) relation \citep{Kennicutt1998} to infer the gas surface density ($\Sigma_{\rm gas}$). The KS relation is defined as, 
\begin{equation}
\rm \left(\frac{\Sigma_{SFR}}{M_{\odot}yr^{-1}kpc^{-2}}\right)\,=\,A \left(\frac{\Sigma_{gas}}{M_{\odot}pc^{-2}}\right)^n,
\end{equation}
\noindent where A=1.5\,$\times$\,10$^{-4}$\,M$_{\odot}$yr$^{-1}$pc$^{-2}$ and $n$\,=\,1.4. Galaxies with Q$_{\rm gas}$\,<\,1 are unstable to local gravitational collapse and will fragment into clumps. Galaxies with  Q$_{\rm gas}$\,>\,1 have sufficient rotational support for the gas and are stable against collapse. We are assuming that the galaxy averaged Q$_{\rm gas}$ is a good approximation of the average disc stability as we do not spatially resolve Q$_{\rm gas}$. We note however that we are primarily using Q$_{\rm gas}$ to differentiate across the KGES sample, and it is the relative value of Q$_{\rm gas}$ that is important rather than focusing on the specific stability of each galaxy. We also note that this parameter only describes the stability of a pure gas disc. The stability of a disc composed of gas and stars is given by the total Toomre Q$_{\rm t}$\,$\approx$\,1/(1/Q$_{\rm gas}$+1/Q$_{\rm stars}$) and describes stability against Jeans clumps. For a more in-depth analysis of the relation between Toomre Q and galaxy properties see \citet{Romeo2018}.

We measure the Toomre Q$_{\rm gas}$ parameter in all 243 KGES galaxies identifying a median stability parameter of $\langle$\,Q$_{\rm gas}$\,$\rangle$\,=\,0.63\,$\pm$\,0.10. We note this is not the true value of disc stability for the KGES sample since we do not take into account the disc thickness nor the stability of the stellar component \citep[e.g][]{Wang1994,Romeo2011}


To understand the dynamical differences between galaxies of different morphologies, we separate out the spheroidal, disc and peculiar galaxies 
and study their dynamical and morphological properties. 
In Figure \ref{Fig:morph_hist} we show example $HST$ colour images of spheroidal, disc and peculiar galaxies in the KGES sample, as well as the distributions of various morphological and kinematic parameters. In comparison to the disc galaxies in the KGES sample, spheroidal galaxies on average have lower specifc angular momentum and are more dispersion dominated but have velocity dispersions that are comparable: $\langle$\,$\sigma_{\rm 0,\, spheroidal}$\,$\rangle$\,=\,56\,$\pm$\,9\,km\,s$^{-1}$ and $\langle$\,$\sigma_{\rm 0,\, disc}$\,$\rangle$\,=\,58\,$\pm$\,6 \,km\,s$^{-1}$. The spheroidal galaxies are more unstable to local gravitational collapse  with higher H$\alpha$ star formation rate surface densities, where $\langle$\,$\Sigma_{\rm SFR,\, disc}$\,$\rangle$\,=\,0.09\,$\pm$\,0.04 M$\rm_{\odot}yr^{-1}kpc^{-2}$ compared to $\langle$\,$\Sigma_{\rm SFR,\, spheroidal}$\,$\rangle$\,=\,0.77\,$\pm$\,0.21 M$\rm_{\odot}yr^{-1}kpc^{-2}$. Morphologically they are less clumpy and more concentrated, but have very similar asymmetries with $\langle$\,A$_{\rm spheroidal}$\,$\rangle$\,=\,0.19\,$\pm$\,0.04 and $\langle$\,A$_{\rm disc}$\,$\rangle$\,=\,0.19\,$\pm$\,0.03.

Taking the properties of morphologically peculiar galaxies in the KGES sample in comparison to morphologically disc dominated galaxies, we establish that on average they have comparable levels of specific angular momentum, velocity dispersion and are equally rotation dominated with $\langle$\,V(2R$_{\rm h}$)/$\sigma_0$\,$_{\rm disc}$\,$\rangle$\,=\,2.33\,$\pm$\,0.40 and $\langle$\,V(2R$_{\rm h}$)/$\sigma_0$\,$_{\rm peculiar}$\,$\rangle$\,=\,2.22\,$\pm$\,0.37. A peculiar galaxy has comparable stability against gravitational collapse to a disc galaxy, with higher $\Sigma_{\rm SFR}$ where $\langle$\,$\Sigma_{\rm SFR,\, peculiar}$\,$\rangle$\,=\,0.25\,$\pm$\,0.08 M$\rm_{\odot}yr^{-1}kpc^{-2}$. Morphologically peculiar galaxies are more clumpy and asymmetrical with slightly higher levels of concentration with $\langle$\,C$_{\rm peculiar}$\,$\rangle$\,=\,2.33\,$\pm$\,0.09 whilst $\langle$\,C$_{\rm disc}$\,$\rangle$\,=\,2.38\,$\pm$\,0.12.

\subsubsection{Interpretation - The High-Redshift Galaxy Demographic}

From Figure \ref{Fig:morph_hist}, for a given stellar mass, a galaxy with low specific angular momentum is likely to be spheroidal, whilst a galaxy with high specific angular momentum and high star formation rate surface density is likely to be peculiar. High specific angular momentum galaxies with low star formation rate surface density, on average, tend to have disc-like morphologies.

Assuming the  galaxies in the KGES sample follow the Kennicutt-Schmidt relation \citep[e.g][]{Gnedin2010,Freundlich2013,Orr2018,Sharda2018}, galaxies with higher star formation rate surface densities, imply higher gas surface densities and hence likely high gas fractions. Recent hydrodynamical zoom-in simulations with the FIRE project (\citealt{Hopkins2014, Hopkins2018}), have shown that the stellar morphology and kinematics of Milky Way mass galaxies correlate more strongly with the gaseous histories of the galaxies \citep{Garrison-Kimmel2018}, in particular around the epoch the galaxy has formed half of its stars (e.g. $z$\,$\sim$\,1.5 \citealt{Gillman2019}). This indicates the gas content of high--redshift galaxies plays a crucial in the their evolution. The balance between the self-gravity of the gas clouds and the shear due to the galaxy's differential rotation, determines the local gravitational stability of the disc.

Figure \ref{Fig:morph_hist} indicates that peculiar galaxies on average are as stable as disc systems with $\langle$\,Q$_{\rm g,\, disc}$\,$\rangle$\,=\,0.70\,$\pm$\,0.20 whilst $\langle$\,Q$_{\rm g,\, peculiar}$\,$\rangle$\,=\,0.64\,$\pm$\,0.13, but have similar velocity dispersions. Peculiar systems have higher star formation rate surface density, thus given that Toomre Q$_{\rm g}$\,$\propto$\,$\kappa$/$\Sigma_{\rm SFR}$, we would expect a `stable' peculiar galaxy to have a higher $\kappa$ value.



We measure the outer gradient of each galaxy's H$\alpha$ rotation curve in the KGES sample, between r\,=\,R$_{\rm h}$ and r\,=\,2R$_{\rm h}$ as a proxy for the $\kappa$ value, given that Toomre Q$_{\rm g}$ is normally measured radially. In this radial range the impact of beam smearing on the rotation curve is reduced compared to the central regions. It has been shown that the shape of a galaxy's rotation curve is strongly correlated with the morphology of a galaxy at $z$\,=\,0 (e.g. \citealt{Sofue2001}), with galaxies of different Hubble--type morphologies from Sa to Sd having characteristically different rotation curves, that reflect the gravitational potential of the galaxy. 


Peculiar galaxies have a median gradient of $\langle$\,$\frac{\delta v_{\rm H\alpha}}{\delta r}$|$_{\rm r\,=\,R_{\rm h}\,-\,2R_{\rm h}}$\,$\rangle$\,=\,3\,$\pm$\,2\,km\,s$^{-1}$\,kpc$^{-1}$ whilst disc galaxies have $\langle$\,$\frac{\delta v_{\rm H\alpha}}{\delta r}$|$_{\rm r\,=\,R_{\rm h}\,-\,2R_{\rm h}}$\,$\rangle$\,=\,4\,$\pm$\,2\,km\,s$^{-1}$\,kpc$^{-1}$. 
The outer gradients of the peculiar galaxies in the KGES sample, at a fixed mass, are very similar to that of disc galaxies, which is reflected in their lower Toomre Q$_{\rm g}$.
This suggests at a fixed stellar mass, high redshift peculiar galaxies are dynamically differentiated from disc dominated galaxies, by their higher $\Sigma_{\rm SFR}$ and higher gas fractions. The peculiar galaxies on average have similar specific angular momentum to disc galaxies, so to evolve to a well ordered Hubble--type galaxies, they do not require additional angular momentum. We predict that through the consumption of their large gas reservoir, via the on-going high levels of star formation, and the fragmentation of the clumpy H{\sc{ii}} regions, driven by the evolution in the characteristic star-forming clump mass \citep[e.g.][]{Livermore2012,Livermore2015}, the angular momentum of the galaxy is re-distributed and the peculiar galaxies evolve to more stable and ordered Hubble-type morphologies.

We note that one possible origin for the peculiar morphology of high redshift galaxies is galaxy interactions which disrupt the steady state dynamics and morphology of galaxies. Galaxy interactions and mergers are much more common in the distant Universe \citep{Rodrigues2017} and would result in increased scatter in the j$_{*}$\,--\,M$_{*}$ plane, depending on the magnitude of the merger and the gas fractions of the galaxies involved. We anticipate only the presence of extremely late state mergers in the KGES sample given the relatively small KMOS field of view and that we identify peculiar and disc galaxies to have comparable specific angular momentum and levels of turbulence.




\section{Conclusions}\label{Sec:Conc}

We have analysed the distribution and correlations of the specific stellar angular momentum (j$_{*}$) in typical $z$\,$\sim$\,1.5 star--forming galaxies by exploiting KMOS H$\alpha$ observations of 288 galaxies from the KGES Survey (Tiley et. al. in prep.). The survey samples the star-formation main-sequence with a broad range of stellar masses, from $\log$(M$_{*}$[M$_{\odot}$])=8.9\,--\,11.7  and H$\alpha$ star-formation rates, with the sample having a 16-84th percentile range of range of SFR\,=\,3\,--\,44\,M$_{\odot}$yr$^{-1}$. We summarise our findings as follows:

\begin{itemize}
\item We use {\sc{galfit}} to measure the structural properties for all 288 galaxies in the KGES survey from $HST$ CANDELS (173 galaxies), archival (96 galaxies) and ground based imaging (19 galaxies). We derive a median half-light radius of $\langle$\,R$_{\rm h}$\,$\rangle$\,=\,0\farcs{31}$\, \pm$\,0\farcs{02} (2.60\,$\pm$\,0.15\,kpc at  $z$\,=\,1.5). We show that KGES galaxies occupy a similar parameter space to typical main--sequence galaxies in the stellar mass--stellar continuum half-light radius plane (Figure \ref{Fig:MS}). 
\item We measure the CAS (Concentration, Asymmetry and Clumpiness) parameters of the galaxies in the KGES survey (Figure \ref{Fig:CAS_Hist}) 
establishing a medium Clumpiness of $\langle$\,S\,$\rangle$\,=\,0.37\,$\pm$\,0.10, Asymmetry of $\langle$\,A\,$\rangle$\,=\,0.19\,$\pm$\,0.05  and a medium Concentration of $\langle$\,C\,$\rangle$\,=\,2.36\,$\pm$\,0.34.  This is similar to the concentration and asymmetry parameters derived for typical main--sequence star--forming galaxies from $z$\,=\,1.5\,--\,3.6 by \citet{Law2012a} with A\,$\sim$\,0.25 and C\,$\sim$\,3. 
\item Taking advantage of the resolved dynamics for 235 galaxies in the sample, we derive the intrinsic H$\alpha$ rotation velocity of each galaxy. We combine the asymptotic rotation velocity and size to measure the 
specific stellar angular momentum and constrain the j$_*$\,--\,M$_*$ plane for the KGES survey (Figure \ref{Fig:JM_Clump}). We quantify the plane with a function of the form  $\log_{10}(j_*)$\,=\,2.61\,+$\,0.66\,(\log_{10}(\rm M_*/M_{\odot})-10.10)$. The normalisation ($\alpha$\,=\,2.61) of this plane is lower than that of $z$\,$\sim$\,0 disc galaxies presented in \citet{Romanowsky2012}
\item To quantify a galaxy's position in the j$_*$\,--\,M$_*$ plane we define a new parameter ($\Delta$j) that is the residual of the logarithm of a galaxy's specific stellar angular momentum and the logarithm of the specific stellar angular momentum of the parametric fit at the same stellar mass. We explore correlations between $\Delta$j and a galaxy's velocity dispersion ($\sigma_0$), establishing no correlation, as well with the  ratio of rotation velocity to velocity dispersion (V(r=2R$_{\rm h}$/$\sigma_0$)) and H$\alpha$ star formation rate surface density ($\Sigma_{\rm SFR}$, Figure \ref{Fig:deltaj_dyn}).
\item Galaxies with higher $\Sigma_{\rm SFR}$, tend to be more dispersion dominated and have lower angular momentum together with visual morphologies resembling spheroidal systems. Rotation dominated galaxies, with low $\Sigma_{\rm SFR}$, have higher angular momentum and have morphologies that resemble discs or peculiar systems.
\item To understand the connection between a galaxy's morphology and specific stellar angular momentum, we take advantage of the multi-band $HST$ CANDELS imaging and derive WFC colour images. In Figure \ref{Fig:JM_Class} we show the j$_*$\,--\,M$_*$ plane coloured by Hubble morphology. We identify a trend of spheroidal galaxies having low angular momentum whilst the more `discy' late-type morphology galaxies have higher angular momentum.
\item We explore the correlation between $\Delta$j and a galaxies parameterised morphology, establishing that higher S\'ersic index, higher stellar bulge to total ratio, galaxies have lower angular momentum, whilst higher angular momentum galaxies have more clumpy morphologies. We propose a picture whereby at a fixed stellar mass spheroidal galaxies have lower angular momentum and are smooth and more symmetrical. Peculiar and disc-like galaxies have higher angular momentum and are much more clumpy.
\item We differentiate peculiar galaxies from disc domianted systems at a fixed stellar mass by analysing their dynamical properties (Figure \ref{Fig:morph_hist}). We derive a median Toomre Q$_{\rm gas}$ of $\langle$\,Q$_{\rm gas}$\,$\rangle$\,=\,0.66\,$\pm$\,0.01 for all 243 KGES galaxies. Peculiar galaxies have higher $\Sigma_{\rm SFR}$, and thus imply higher gas fractions than disc galaxies.
\end{itemize}

Overall, we have identified that the morphologies of high--redshift star--forming galaxies are more complicated than those in the local Universe, but can be split into three broad classes of spheroidal, disc and peculiar. We can dynamically differentiate the three classes at fixed stellar mass, whereby spheroidal galaxies have lower specific angular momentum and high gas fractions, whilst disc-like galaxies have high specific angular momentum and lower gas fractions. Peculiar systems have equally high levels of specific angular momentum as disc galaxies, but have higher gas fractions.

In order to further explore these correlations and establish empirical constraints on how the gas fractions, stellar population demographic and rotation curve gradients define the emergence of peculiar gas rich systems, as well as Hubble-type spirals, we require accurate measurements of gas fractions in these systems e.g. ALMA molecular gas observations, as well as constraints on the metallicity and stellar age of galaxies from multi-line emission line diagnostics. 

\section*{Acknowledgements}

This work was supported by the Science and Technology Facilities 
Council (ST/L00075X/1). SG acknowledges the support of the Science and Technology Facilities Council through grant ST/N50404X/1 for support. E.I.\ acknowledges partial support from FONDECYT through grant N$^\circ$\,1171710. We thank the FMOS-COSMOS team for their invaluable contributions to the KGES target selection. ALT acknowledges support from STFC (ST/L00075X/1 and ST/P000541/1), ERC Advanced Grant DUSTYGAL (321334), and a Forrest Research Foundation Fellowship. LC is the recipient of an Australian Research Council Future Fellowship (FT180100066) funded by the Australian Government. Parts of this research were conducted by the Australian Research Council Centre of Excellence for All Sky Astrophysics in 3 Dimensions (ASTRO 3D), through project number CE170100013.  

The KMOS data in this paper were obtained at the Very Large Telescope of the European Southern Observatory, Paranal, Chile (ESO Programme IDs 095.A-0748, 096.A-0200, 097.A-0182, 098.A-0311, and 0100.A-0134).  This work is
based in part on data obtained as part of the UKIRT Infrared
Deep Sky Survey. This work is based on observations taken by the
CANDELS Multi-Cycle Treasury Program with the NASA/ESA
HST, which is operated by the Association of Universities for
Research in Astronomy, Inc., under NASA contract NAS5-26555.
HST data were also obtained from the data archive at the Space
Telescope Science Institute. 




\bibliographystyle{mnras}
\bibliography{master}

\bsp	



\appendix

\begin{figure*}
\section{SEDs, HST Imaging and Kinematics}\label{App:SEDs}
	\includegraphics[width=0.98\linewidth,,trim={0cm 5cm 0cm 0cm}]{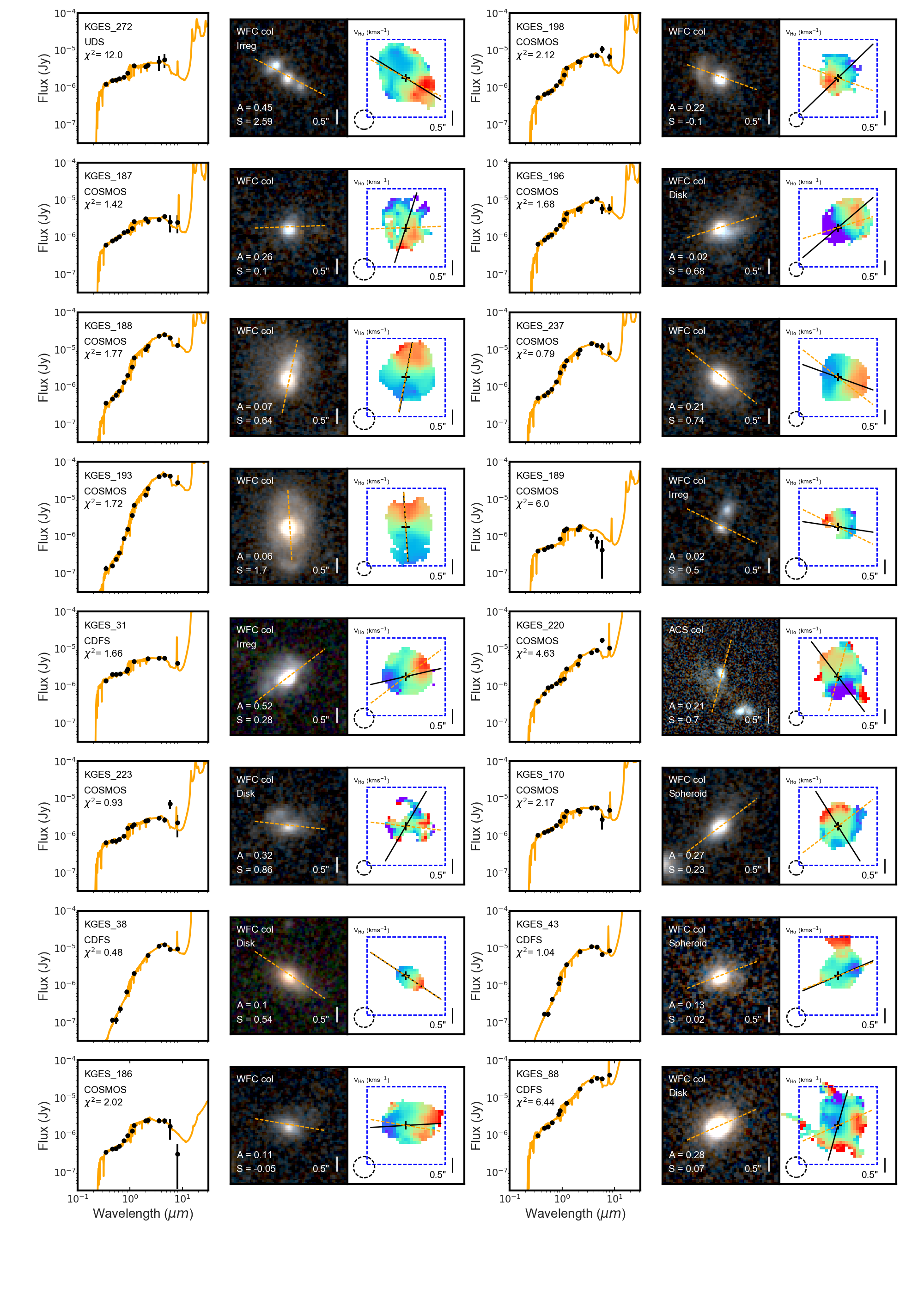}
	\caption{For each galaxy we show the multi-wavelength photometry from  $UV$\,--\,8\,$\micron$ with the derived {\sc{magphys}} SEDs fits (left), the `best' broadband image with semi-major axis (orange line) and asymmetry and clumpiness values stated (middle) and the H$\alpha$ velocity map of the galaxy (right) with kinematic position angle (black line). The additional 17 pages are shown in supplementary material.}
\label{Fig:SEDs}
\end{figure*}
\FloatBarrier

\begin{figure*}
\section{GALFIT Model Examples}\label{App:Galfit}
\centering
\includegraphics[width=1\linewidth,]{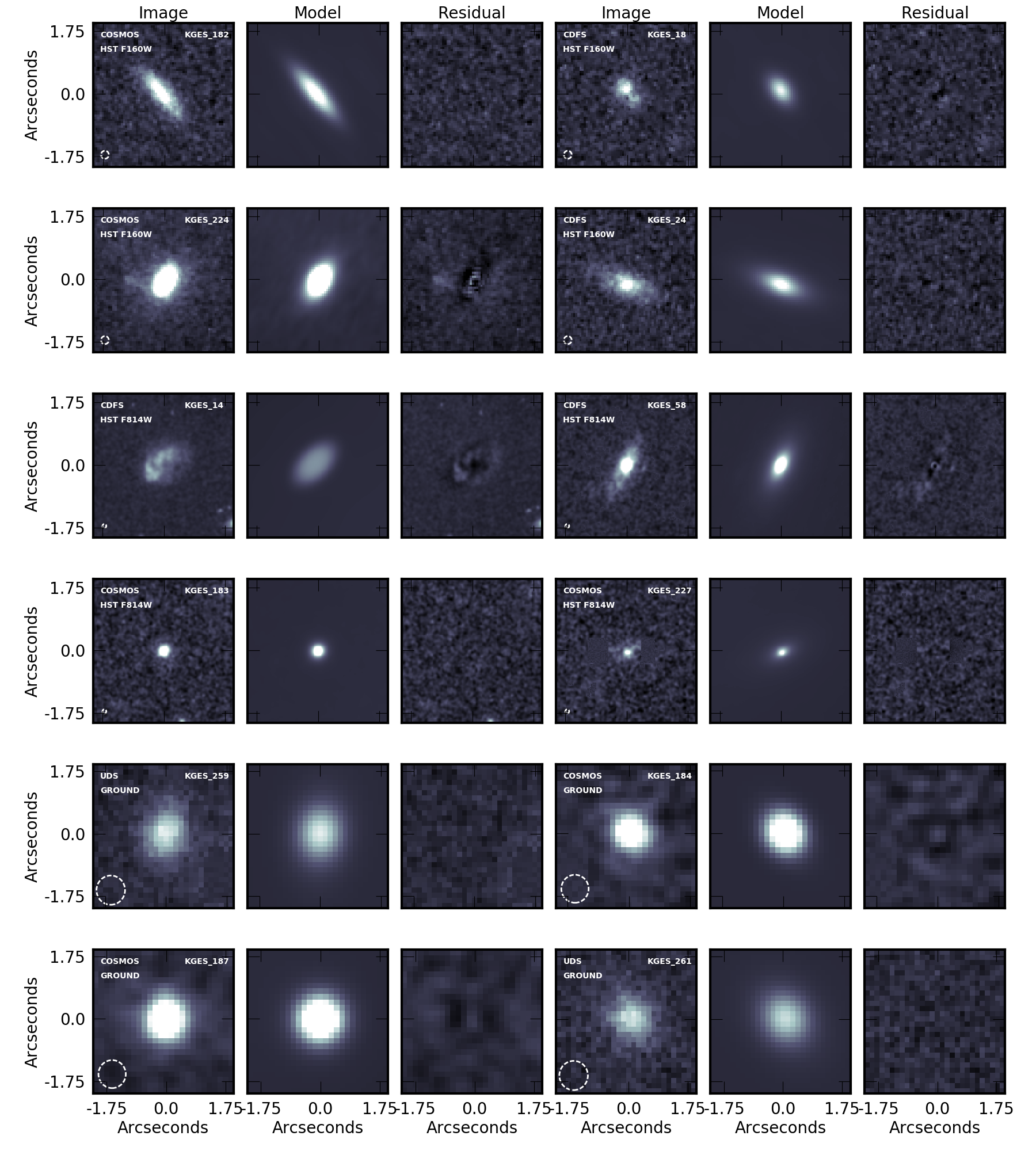}
\caption{Representative 4\,$\times$\,4 arcsecond examples of the imaging, {\sc{galfit}} models and residuals of KGES galaxies from COSMOS, CDFS and UDS extragalatic field in $HST$ F160W, F814W and ground based UKIDDS $K$--band and COSMOS UVISTA $H$\,--\,band images respectively. The PSF of each image is shown by the white circle in the lower left corner of each image. In each case the model recreates the image well and minimises the residual.}
\label{Fig: Galfit}
\end{figure*}
\FloatBarrier


\label{lastpage}
\end{document}